\documentclass[twocolumn]{revtex4}

\usepackage{graphicx}
\usepackage{epsf}
\usepackage{amsmath,amssymb}
\usepackage{ulem}
\usepackage{color}
\newcommand{\bvec}{\boldsymbol}

\newcommand{\AK}{A_K}
\newcommand{\kNkN}{\kbar N+\kbar N}
\newcommand{\barV}{V}
\newcommand{\barT}{{T}^\textrm{kin}_0}

\begin{document}

\newcommand{\kbar}{{\bar K}}

\newcommand{\newlr}{\textrm{LR}}
\newcommand{\newrl}{\textrm{RL}}
\newcommand{\newll}{\textrm{LL}}
\newcommand{\newrr}{\textrm{RR}}

\newcommand{\newrt}{\textrm{R}}
\newcommand{\newlf}{\textrm{L}}
\newcommand{\newlfrt}{\textrm{L,R}}

\title{Binding of antikaons and $\Lambda(1405)$ clusters in light kaonic nuclei}

\author{Yoshiko Kanada-En'yo}
\affiliation{Department of Physics, Kyoto University, Kyoto 606-8502, Japan}

\begin{abstract}
The energy spectra of light-mass kaonic nuclei 
were investigated using the theoretical framework of 
the $0s$-orbital model with
zero-range $\bar{K} N$ and $\bar{K}\bar{K}$ interactions of effective single-channel real potentials.
The energies of the $\bar{K} NN$, $\bar{K} NNN$, $\bar{K} NNNN$, $\bar{K}\bar{K} N$,
and $\bar{K}\bar{K} NN$ systems were calculated in the cases of weak- and deep-binding of the
$\bar{K} N$ interaction, which was adjusted to fit the $\Lambda(1405)$ 
mass with the energy of the $\bar{K} N$ bound state. 
The results qualitatively reproduced the energy systematics of kaonic nuclei calculated via 
other theoretical approaches. 
In the energy spectra of the $\bar{K} NN$ and  $\bar{K}\bar{K} NN$ systems, 
the lowest states  $\bar{K} NN(J^\pi,T=0^-,1/2)$ and $\bar{K}\bar{K} NN(0^+,0)$ were found to have 
binding energies approximately twice and four times as large as that 
of the $\bar{K} N(1/2^-,0)$ state, respectively. 
Higher $(J^\pi,T)$ states including 
$\bar{K} NN(1^-,1/2)$,  $\bar{K}\bar{K} NN(0^+,1)$, and $\bar{K}\bar{K} NN(1^+,1)$ were predicted
at energies of 9--25 MeV below the antikaon-decay threshold.
The effective $\Lambda(1405)$-$\Lambda(1405)$ interaction in the  $\bar{K}\bar{K} NN$ system
was also investigated via a $\bar{K} N+\bar{K} N$-cluster model.
Strong and weak $\Lambda(1405)$-$\Lambda(1405)$ attractions were obtained 
in the $S^\pi=0^+$ and $S^\pi=1^-$ channels, respectively.
The $\Lambda(1405)$-$\Lambda(1405)$ interaction in the $\bar{K}\bar{K} NN$ system was compared 
with the effective  $d$-$d$ interaction in the $NNNN$ system, and the 
properties of dimer-dimer interactions in hadron and nuclear systems were discussed.
\end{abstract}

\maketitle

\section{Introduction}

Kaonic nuclei have recently become a hot topic in hadron physics. 
In particular, light-mass kaonic nuclei have been intensively 
investigated to understand the structures of exotic hadrons that 
are interpreted as multihadron systems called hadron molecular states.
One candidate for the simplest such system 
is the $\Lambda(1405)$ state (denoted as $\Lambda^*$), which is 
the lowest negative-parity $\Lambda$ state with $(J^\pi,T)=(1/2^-,0)$. 
The $\Lambda(1405)$ state is a narrow resonance observed in the $\pi\Sigma$ spectra at an energy 
slightly below the $\kbar N$ threshold, and is considered to be a $\kbar N$ quasibound state 
produced by a strong $\kbar N$ interaction. 
This picture of the $\kbar N$ molecular state for the $\Lambda(1405)$ resonance led to 
the concept of kaonic states with an antikaon deeply bound via the $\kbar N$ interaction
in light-mass nuclei 
such as $K^- pp$ and $K^- ppn$ \cite{Akaishi:2002bg,Yamazaki:2002uh,Dote:2003ac,Akaishi:2005sn,Yamazaki:2007cs},
for which various few-body calculations have been developed \cite{Shevchenko:2006xy,Shevchenko:2007zz,Ikeda:2007nz,Ikeda:2008ub,Ikeda:2010tk,Dote:2008in,Dote:2008hw,Dote:2014via,Wycech:2008wf,Bayar:2011qj,Oset:2012gi,Barnea:2012qa,Maeda:2013zha,Ohnishi:2017uni}.
Several experiments have been 
performed in searching of the $\kbar NN$ state~\cite{Agnello:2005qj,Suzuki:2004ep,Sato:2007sb,Suzuki:2007kn}, but
the evidence has not yet been confirmed~\cite{Magas:2006fn,Yamazaki:2010mu}. 
A similar challenging is investigating double-kaonic nuclei with two antikaons 
\cite{Yamazaki:2003hs,Dote:2005nb,Kanada-Enyo:2008wsu}.

For kaonic nuclei of mass number $A=2$, 
intensive studies of the $\kbar NN$ system have been performed by many groups. 
To clarify the properties of the strangeness dibaryon, the $\kbar\kbar NN$ system is also a key issue. 
It has also attracted interest in properties of the $\kbar\kbar$ interaction 
concerning the kaon condensation in dense nuclear matter.
To experimentally search for the quasibound $\kbar\kbar NN$ state, 
a formation mechanism via a $\Lambda^*+\Lambda^*$ doorway was proposed \cite{Hassanvand:2011zz}.
Furthermore, the effective $\Lambda^*$-$\Lambda^*$ interaction in the $\kbar\kbar NN$ system 
may draw general interest in dimer-dimer interactions in hadron systems. 
For such a system consisting of isospin SU(2) bosons and fermions, the question of  
what role is played by the nucleon Fermi statistics and antikaon Bose statistics in the effective dimer-dimer 
interaction between two $\Lambda^*$ particles arises. 

The original idea for kaonic nuclei was based on the phenomenological $\kbar N$ interaction
called the Akaishi-Yamazaki (AY) interaction \cite{Akaishi:2002bg,Yamazaki:2007cs}. The AY interaction
is characterized by an extremely strong $\kbar N$ attraction in the $T=0$ channel, which reproduces
an $\Lambda(1405)$ mass at an energy 27 MeV below the $\kbar N$ threshold 
reported in Particle Data Group (PDG) table \cite{Tanabashi:2018oca}.
On the other hand, a weaker $\kbar N$ interaction was proposed
in detailed analyses of $\pi\Sigma$ scattering based on the chiral SU(3) effective field theory, 
from which the $\Lambda^*$ resonance-pole position was obtained at only 8--12~MeV below the  $\kbar N$ threshold 
\cite{Hyodo:2007jq}.
In three-body calculations of the $\kbar NN$, the deep-type AY interaction predicted a deeply bound 
$\kbar NN$ state, 
whereas the weak-type chiral interaction obtained a smaller binding energy (B.E.) for the $\kbar NN$ state.
Moreover, for other kaonic nuclei, theoretical predictions of the binding energies are spread over a range
because of the uncertainty of $\kbar N$ interactions 
and model ambiguities in theoretical treatments, such as 
the energy dependence of the interaction as well as channel coupling.

In this paper, my aim is to investigate the energy spectra of single- and double-kaonic nuclei, 
particularly, the $\kbar\kbar NN$ system.
I do not intend to predict precise values of the energy spectra, which may depend upon 
the details of hadron-hadron interactions as well as model treatments.
Instead, I investigate the energy systematics of kaonic nuclei to understand their
global features and to extract universal properties independently from 
the uncertainty in hadron-hadron interactions.
In this paper, I apply a simple model of the $0s$-orbital  configuration 
to kaonic nuclei and calculate their energy spectra by assuming
zero-range $\kbar N$ and $\kbar\kbar$ interactions of effective single-channel real potentials.
The $\kbar N$ interaction is tuned to fit the $\Lambda^*$ mass with the $\kbar N$ bound state 
in two cases of weak- and deep-binding. 
As for the $NN$ interaction, 
I adopt a finite-range effective $NN$ interaction adjusted to reproduce $S$-wave $NN$-scattering lengths.
I discuss the important roles of isospin symmetry
in the energy spectra of kaonic nuclei.
I also investigate the effective $\Lambda^*$-$\Lambda^*$ interaction with a 
$\kbar N+\kbar N$-cluster model. For comparison with the effective $d$-$d$ interaction 
in the $NNNN$ system, I discuss the properties of 
dimer-dimer interactions and binding mechanisms in hadron and nuclear systems.

This paper is organized as follows. In Sec.~\ref{sec:formulation}, the theoretical frameworks
for the $0s$-orbital and $\kbar N+\kbar N$ cluster models
are explained. In Sec.~\ref{sec:results-1}, the results of the $0s$-orbital model for kaonic nuclei 
are presented. 
In Sec.~\ref{sec:results-2}, 
the $\kbar \kbar NN$ system is
investigated via the $\kbar N+\kbar N$-cluster model, 
and the effective $\Lambda^*$-$\Lambda^*$ interactions are discussed. Finally, 
a summary is given in Sec.~\ref{sec:summary}.

\section{Formulation of single- and double-kaonic nuclei}\label{sec:formulation}
\subsection{$0s$-orbital model for single- and double-kaonic nuclei}
For single and double-kaonic nuclei
with $\AK$ antikaons and $A$ nucleons, 
I assume the $0s$-orbital  configuration
with one-range 
Gaussian wave functions and express the wave functions for the $(J^\pi,T)$ states with 
angular momentum $(J)$, parity $(\pi)$, and isospin $(T)$ as
\begin{align}
&\Psi^{(J^\pi, T)}_{\kbar^{\AK} N^{A}}(1',\ldots,\AK',1,\ldots,A)\nonumber\\
& =
\phi^K_{0}({\bvec{r}}_{1'})\cdots
\phi^K_{0}(\bvec{r}_{\AK'})
\phi^N_{0}(\bvec{r}_1)\cdots
\phi^N_{0}(\bvec{r}_{A})\nonumber\\
&\quad \otimes
[s_1\cdots s_A]_S \otimes [ t_{1'}\cdots t_{\AK'}t_1 \cdots t_A]_{T},\\
&\phi^N_0({\bvec{r}})=
 \left(\frac{2\nu_N}{\pi}\right)^{3/4}
e^{-\nu_N{\bvec{r}}^2},\\
&\phi^K_0({\bvec{r}})=
 \left(\frac{2\nu_K}{\pi}\right)^{3/4}
e^{-\nu_K{\bvec{r}}^2}
\end{align}
with $J=S$ and $\pi=(-1)^{A_K}$.  
Here, nucleon spins $s_i$ are coupled to the total nuclear spin $S$, and 
antikaon isospins $t_{i'}$ and nucleon isospins $t_i$ are coupled to the total isospin $T$. 
The spatial configuration of identical particles in the $0s$-orbit
is symmetric. 
The nucleon Fermi statistics are
taken into account in the nucleon-spin and -isospin configuration. 
For double-kaonic nuclei,
the isospins of two antikaons in the $0s$-orbit 
are coupled to form an isovector ($\tau_K=1$) as $[t_{1'}t_{2'}]_{\tau_K=1}$ 
 to satisfy the Bose statistics.

The Gaussian width parameter $\nu_K$ for $\phi^K_{0}$ is chosen to be 
\begin{align}
\nu_K\equiv \frac{m_K}{m_N}\nu_N
\end{align}
with a ratio $m_K/m_N\approx 1/2$ of the antikaon mass $m_K$ to the nucleon mass $m_N$, such that 
the center-of-mass (cm) motion can be exactly removed from the total wave function.
Hence, the antikaon orbit $\phi^K_{0}$ has a broader distribution than the 
nucleon orbit $\phi^N_{0}$. The internal wave functions 
of the $NN$, $\kbar \kbar$, and $\kbar N$ pairs can be written as
\begin{align}
&\Phi^{\kbar\kbar}(\bvec{r}_{i'j'})= \left ( \frac{\nu_K}{\pi}\right )^{3/4}
e^{-\frac{\nu_K}{2}\bvec{r}^2_{i'j'}}, \\
&\Phi^{NN}(\bvec{r}_{ij})= \left ( \frac{\nu_N}{\pi}\right )^{3/4}
e^{-\frac{\nu_N}{2}\bvec{r}^2_{ij}}, \\
&\Phi^{\kbar N}(\bvec{r}_{i'j})= \left ( \frac{\lambda}{\pi}\right )^{3/4}
e^{-\frac{\lambda}{2}\bvec{r}_{i'j}^2},\\
&\lambda=\frac{2 m_\kbar}{m_N+m_\kbar}\nu_N,
\end{align}
where $\bvec{r}_{kl}=\bvec{r}_l-\bvec{r}_k$.
The root-mean-square (rms) distances $\sqrt{\langle r^2_{kl} \rangle}$ 
of the $NN$, $\kbar\kbar$, and $\kbar N$ pairs are given as $R_{NN}=1/\sqrt{2\nu_N}$, 
$R_{\kbar\kbar}=1/\sqrt{2\nu_K},$ and 
$R_{\kbar N}=1/\sqrt{2\lambda}$, respectively.

\subsubsection{Wave functions of single-kaonic nuclei}
The $\kbar N$ bound state 
with 
$(J^\pi, T)=(\frac{1}{2}^-,0)$ corresponding to $\Lambda^*$ is expressed as  
\begin{align}\label{eq:KN-wf}
&\Phi^{(\frac{1}{2}^-,0)}_{\kbar N}(1',1)=
\phi^K_{0}({\bvec{r}_{1'}})\phi^N_{0}({\bvec{r}}_1)\otimes
s_1 \otimes [t_{1'} t_1]_{T=0}, 
\end{align}
where the nucleon spin $s_1=\{\uparrow, \downarrow\}$ specifies the intrinsic spin of the $\Lambda^*$ system.

The $\kbar NN$ states with $(J^\pi, T)=(0^-,\frac{1}{2})$, 
$(0^-,\frac{3}{2})$, and $(1^-,\frac{1}{2})$ are written as 
\begin{align}
\Psi^{(0^-, T)}_{\kbar NN}(1',1,2) &=
\phi^K_{0}({\bvec{r}}_{1'})
\phi^N_{0}(\bvec{r}_1)
\phi^N_{0}(\bvec{r}_{2})\nonumber\\
& \otimes
[s_1 s_2]_{S=0} \otimes \Bigl[t_{1'}[t_1 t_2]_{\tau_N=1}\Bigr]_{T},\\
\Psi^{(1^-, \frac{1}{2})}_{\kbar NN}(1',1,2) &=
\phi^K_{0}({\bvec{r}}_{1'})
\phi^N_{0}(\bvec{r}_1)
\phi^N_{0}(\bvec{r}_{2})\nonumber\\
& \otimes
[s_1 s_2]_{S=1} \otimes \Bigl[t_{1'}[t_1 t_2]_{\tau_N=0}\Bigr]_{T=\frac{1}{2}},
\end{align}
where $\tau_N$ indicates the total nucleon isospin.
The $J=0$ states contain an isovector $NN$ pair, and the 
$J=1$ state contains a deuteron-like isoscalar $NN$ pair
because of the nucleon Fermi statistics for the total nucleon spin, $S=J$. 
The $(0^-,\frac{1}{2})$ state is the lowest 
$\kbar NN$ state, which has been intensively studied by three-body calculations,
whereas the $(1^-,\frac{1}{2})$ state was predicted to be 
a higher $\kbar NN$ state~\cite{Bayar:2011qj,Oset:2012gi}.

For kaonic nuclei with mass numbers $A=3$ and $A=4$, 
I consider the $(J^\pi, T)=(\frac{1}{2}^-, 0)$ and $(0^-,\frac{1}{2})$ 
states with $^3$H and $^4$He cores, respectively, as 
\begin{align}
&\Psi^{(\frac{1}{2}^-, 0)}_{\kbar NNN}(1',1,2,3) =
\phi^K_{0}({\bvec{r}}_{1'})
\phi^N_{0}(\bvec{r}_1)
\phi^N_{0}(\bvec{r}_{2})\phi^N_{0}(\bvec{r}_{3})\nonumber\\
& \qquad\qquad \otimes
[s_1 s_2 s_3]_{S=\frac{1}{2}} \otimes \Bigl[t_{1'}[t_1 t_2 t_3]_{\tau_N=\frac{1}{2}}\Bigr]_{T=0},\\
&\Psi^{(0^-,\frac{1}{2})}_{\kbar NNNN}(1',1,2,3,4) \nonumber\\
&\qquad =
\phi^K_{0}({\bvec{r}}_{1'})
\phi^N_{0}(\bvec{r}_1)
\phi^N_{0}(\bvec{r}_{2})\phi^N_{0}(\bvec{r}_{3})\phi^N_{0}(\bvec{r}_{4})\nonumber\\
&\qquad \otimes
[s_1 s_2 s_3 s_4]_{S=0} \otimes \Bigl[t_{1'}[t_1 t_2 t_3 t_4 ]_{\tau_N=0}\Bigr]_{T=\frac{1}{2}}.
\end{align}

\subsubsection{Wave functions of double-kaonic nuclei}

The $0s$-orbital states of double-kaonic nuclei contain an isovector $\kbar\kbar$ pair because of the Bose statistics.
The $\kbar\kbar N$ system with $(J^\pi, T)=(\frac{1}{2}^+, 0)$
is given by
\begin{align}
\Psi^{(\frac{1}{2}^+, \frac{1}{2})}_{\kbar\kbar N}(1',2',1) &=
\phi^K_{0}({\bvec{r}}_{1'})\phi^K_{0}({\bvec{r}}_{2'})
\phi^N_{0}(\bvec{r}_1)\nonumber\\
&  \otimes
s_1\otimes \Bigl[[t_{1'}t_{2'}]_{\tau_K=1} t_1 \Bigr]_{T=\frac{1}{2}}.
\end{align}

For the $\kbar\kbar NN$ system with the $0s$-orbital  configuration, 
the $(J^\pi, T)=(0^+, T)$ states consist of 
isovector $NN$ and $\kbar\kbar$ pairs, which are coupled to 
the total isospins $T=0$, 1, and 2, and 
the $(J^\pi, T)=(1^+, T)$ state contains an isoscalar $NN$ pair and 
an isovector $\kbar\kbar$ pair 
as 
\begin{align}
&\Psi^{(0^+, T)}_{\kbar\kbar NN}(1',2',1,2) =
\phi^K_{0}({\bvec{r}}_{1'})\phi^K_{0}({\bvec{r}}_{2'})
\phi^N_{0}(\bvec{r}_1)\phi^N_{0}(\bvec{r}_2)\nonumber\\
&\qquad  \otimes
[s_1 s_2]_{S=0} \otimes \Bigl[[t_{1'}t_{2'}]_{\tau_K=1}[t_{1}t_{2}]_{\tau_N=1}\Bigr]_{T},\\
&\Psi^{(1^+, 0)}_{\kbar\kbar NN}(1',2',1,2) =
\phi^K_{0}({\bvec{r}}_{1'})\phi^K_{0}({\bvec{r}}_{2'})
\phi^N_{0}(\bvec{r}_1)\phi^N_{0}(\bvec{r}_2)\nonumber\\
&\qquad   \otimes
[s_1 s_2]_{S=1} \otimes \Bigl[[t_{1'}t_{2'}]_{\tau_K=1}[t_{1}t_{2}]_{\tau_N=0}\Bigr]_{T=1}.
\end{align}

\subsection{$\kNkN$ cluster model for $\kbar\kbar NN$}\label{sec:NK-NK-cluster}

To discuss the effective  $\Lambda^*$-$\Lambda^*$ interaction, 
I apply a $\kNkN$-cluster model to the $\kbar\kbar NN$ system. 
I consider two ($\kbar N$)-clusters with the $0s$-orbital  configuration located at 
$-\bvec{R}/2$ and $\bvec{R}/2$
with a distance of $R=|\bvec{R}|$ as 
\begin{align}
&\Psi^{(S^\pi T)}_{\kNkN}(\bvec{R};1',2',1,2)\nonumber\\
&=n_0 {\cal A}_{12} {\cal S}_{1'2'} 
\Bigl\{\phi^K_{-\frac{\bvec{R}}{2}} ({\bvec{r}_{1'}})\phi^N_{-\frac{\bvec{R}}{2}} ({\bvec{r}_1})
\phi^K_{\frac{\bvec{R}}{2}} ({\bvec{r}_{2'}})\phi^N_{\frac{\bvec{R}}{2}} ({\bvec{r}_2})
\nonumber\\
&\otimes
[s_1s_{2}]_S \otimes \Bigr[[t_{1'}t_{2'}]_{\tau_K}[t_{1}t_{2}]_{\tau_N} \Bigr]_{T} \Bigr\},\\
&\phi^K_{\bvec{X}}(\bvec{r})= \left(\frac{2\nu_K}{\pi}\right)^{3/4}
e^{-\nu_K(\bvec{r}-\bvec{X})^2},\\
&\phi^N_{\bvec{X}}(\bvec{r})= \left(\frac{2\nu_N}{\pi}\right)^{3/4}
e^{-\nu_N(\bvec{r}-\bvec{X})^2},
\end{align}
where $n_0$ is the normalization factor. The operators ${\cal A}_{12}$ and ${\cal S}_{1'2'}$ are the
antisymmetrized and symmetrized operators for nucleons and kaons, respectively,
\begin{align}
{\cal A}_{12}=\frac{1-P_{12}}{\sqrt{2}}, \qquad {\cal S}_{1'2'}=\frac{1+P_{1'2'}}{\sqrt{2}},
\end{align}
which are equivalent to the internal-parity-projection operators of the $NN$ and $\kbar\kbar$ pairs.

Hereafter, I consider the $T=0$ states with $\tau_K=\tau_N\equiv\tau$ leading 
to the asymptotic $\Lambda^*+\Lambda^*$ state at $R\to \infty$,
and take the isospin $\tau=0$ and $\tau=1$ components, which I denoted 
as $\Psi^{(S^\pi 0)}_{\kNkN}(\bvec{R},\tau;1',2',1,2)$, into account. 
The wave function with configuration mixing ($\tau$-mixing) of $\tau=0$ and $\tau=1$ 
is given by
\begin{align}
&\Psi^{(S^\pi 0)}_{\kNkN}(\bvec{R};1',2',1,2)\nonumber\\
& =
\sum_{\tau=0,1} c_\tau  \Psi^{(S^\pi 0)}_{\kNkN}(\bvec{R},\tau;1',2',1,2), 
\end{align}
where the coefficients $c_\tau$ are determined by  diagonalization of the norm and Hamiltonian
matrices for $\tau=\{1,0\}$ at each distance $R$. 
Note that the parity $\pi$ of the $\kNkN$ system is related to the total nucleon spin $S$ as $S^\pi=0^+$ and $1^-$
because of the nucleon Fermi and antikaon Bose statistics. These correspond to 
the selection rule of $S^\pi=0^+$ and $1^-$ for two $\Lambda^*$ particles in Fermi statistics.

For the $\kNkN$-cluster state in the $S^\pi=0^+$ channel,  
the $\tau=1$ component has a spatial-even $\kbar\kbar$ pair and a singlet-even $NN$ pair, while
the $\tau=0$ component consists of a spatial-odd $\kbar\kbar$ pair and a singlet-odd $NN$ pair. 
The former, the $\tau=1$ component, becomes equivalent to the lowest  
$0s$-orbital $\kbar\kbar NN$ state with $(J^\pi,T)=(0^+,0)$ at $R=0$. The latter 
is forbidden in the $0s$-orbital model space and therefore goes to an excited configuration with 
two $0p$-orbital particles in the $R\to 0$ limit.

The $\kNkN$-cluster state in the $S^\pi=1^-$ channel leads 
to a spin-aligned $\Lambda^*+\Lambda^*$ state having negative parity. 
The $\tau=1$ component of the $S^\pi=1^-$ state
 has a spatial-even $\kbar\kbar$ pair and a triplet-odd  $NN$ pair, 
whereas the $\tau=0$ component is composed of 
a spatial-odd $\kbar\kbar$ pair and a triplet-even  $NN$ pair. 
Neither component is not allowed in the $0s$-orbital  configuration. Instead, the $\tau=1$ component 
becomes a one-antikaon excitation and the $\tau=0$ component 
goes to a one-nucleon excitation in the $R\to 0$ limit.

It is worth noting that 
the spatial part of the $\kNkN$-cluster wave functions before ${\cal A}_{12}$ and ${\cal S}_{1'2'}$
can be rewritten 
in the separable form of 
the cm, inter-cluster, and $\kbar N$-cluster internal wave functions
with Jacobi coordinates 
as 
\begin{align}
&\phi^K_{-\frac{\bvec{R}}{2}} ({\bvec{r}_{1'}})\phi^N_{-\frac{\bvec{R}}{2}} ({\bvec{r}_1})
\phi^K_{\frac{\bvec{R}}{2}} ({\bvec{r}_{2'}})\phi^N_{\frac{\bvec{R}}{2}} ({\bvec{r}_{2}})\nonumber\\
&=\phi_\textrm{cm}(\bvec{r}_\textrm{cm})\otimes\phi_\textrm{rel}(\bvec{R},\bvec{r}_\textrm{rel})
\otimes\Phi^{\kbar N}(\bvec{r}_{1'1})\otimes\Phi^{\kbar N}(\bvec{r}_{2'2}),\\
&\phi_\textrm{cm}(\bvec{r}_\textrm{cm})=\left( \frac{2\cdot 4\gamma}{\pi} \right) e^{-4\gamma \bvec{r}_\textrm{cm}^2},\\
&\phi_\textrm{rel}(\bvec{R}, \bvec{r}_\textrm{rel})= \left ( \frac{2\gamma}{\pi}\right )^{3/4}
e^{-\gamma(\bvec{r}_\textrm{rel}-\bvec{R})^2},\\
&\gamma=\frac{m_1+m_2}{2m_1}\nu_N,\\
&\bvec{r}_\textrm{cm}=\frac{1}{2} \Bigl\{\frac{m_N\bvec{r}_{1'}+m_K\bvec{r}_1}{m_N+m_K}
+\frac{m_N\bvec{r}_{2'}+m_K\bvec{r}_2}{m_N+m_K}\Bigr\},\\
&\bvec{r}_\textrm{rel}=\frac{m_N\bvec{r}_{2'}+m_K\bvec{r}_2}{m_N+m_K}
-\frac{m_N\bvec{r}_{1'}+m_K\bvec{r}_1}{m_N+m_K}.
\end{align}


\subsection{Hamiltonian and two-body interactions} 

In the present calculation, I omit the charge-symmetry breaking and the Coulomb force. 
The Hamiltonian for single-kaonic nuclei $\kbar N^A$ with mass number $A$ is given by
\begin{align}
&H_{\kbar N^A}=t^\textrm{kin}_{1'}+ \sum^A_{i} t^\textrm{kin}_i -{T}^\textrm{kin}_\textrm{cm}\nonumber\\
&+\sum_{T=0,1} \sum^A_{i} v^{T}_{\kbar N}(1',i)+
\sum_{(ST)} \sum_{i<j} v^{(ST)}_{NN}(i,j),
\end{align}
where  $t^\textrm{kin}$ is the single-particle kinetic energy, $T^\textrm{kin}_\textrm{cm}$ is the cm kinetic energy, 
$v^{T}_{\kbar N}$ is the $T$ component of the ${\kbar N}$ interaction, and $v^{(ST)}_{NN}$ 
with $(ST)=(10),(01),(11)$, and $(01)$ 
indicate the triplet-even, singlet-even, triplet-odd, and singlet-odd components of the $NN$ 
interaction, respectively.
Similarly, the Hamiltonian for the double-kaonic nuclei $\kbar\kbar N^A$ is given by 
\begin{align}
&H_{\kbar\kbar N^A}=t^\textrm{kin}_{1'}+t^\textrm{kin}_{2'}+ \sum^A_{i} t^\textrm{kin}_i -{T}^\textrm{kin}_\textrm{cm}\nonumber\\
&+\sum_{T=0,1} \sum^A_{i} v^{T}_{\kbar N}(1',i)+\sum_{T=0,1} \sum^A_{i} v^{T}_{\kbar N}(2',i)\nonumber\\
&+\sum_{(ST)} \sum_{i<j} v^{(ST)}_{NN}(i,j)+\sum_{T=0,1} v^{T}_{\kbar \kbar}(1',2').
\end{align}
In the present $0s$-orbital model,
the single-particle kinetic energies of antikaons and nucleons have the same value as
\begin{align}
 \langle \phi^N_0 | t^\textrm{kin} |  \phi^N_0\rangle=\langle \phi^K_0 | t^\textrm{kin} |  \phi^K_0\rangle=\frac{3\hbar\omega}{4}\equiv \barT,
\end{align}
where $\hbar\omega=2\hbar^2\nu_N/m_N$. The cm kinetic energy term is also the same value 
$\langle {T}^\textrm{kin}_\textrm{cm}\rangle=\barT$. Thus, the total kinetic energy is given by
$\langle T^\textrm{kin}\rangle=(A_\textrm{tot}-1)\barT$, where $A_\textrm{tot}$ is the total particle number
$A_\textrm{tot}=A+A_K$.

For the $NN$ interaction, I adopt a finite-range effective central interaction of the Volkov $NN$
force~\cite{Volkov:1965zz}, which is often used with cluster models for nuclear systems. 
In the present calcualtion, the $NN$ spin-orbit and tensor 
interactions are omitted.
The Volkov central $NN$ force is given in two-range Gaussian form as 
\begin{align}
& v^{(ST)}_{NN}(i,j)= u^{(ST)}_{NN}(r_{ij})  P^{(ST)}_{ij}, \\
& u^{(ST)}_{NN}(r) = f^{(ST)}_{NN} \sum_{k=1,2} V_k e^{-\frac{r^2}{\eta_k^2}},
\end{align}
where $P^{(ST)}_{ij}$ is the projection operator to the $(ST)$ state of the 
$NN$ pair. The range parameters $\eta_k$ and global-strength parameters $V_k$
are given in the Volkov parameterization, whereas  
the strength ratios $f^{(ST)}_{NN}$ of four components $(ST)=(10),(01),(11)$, and (00) 
are adjustable parameters, which I tune to fit the $S$-wave $NN$-scattering lengths and 
the $\alpha+\alpha$-scattering phase shifts. 
For the spatial parts of the expectation values of the $NN$ interaction for the $0s$-orbital $NN$ pair, 
I use the notation
\begin{align}
&\barV^{(ST)}_{NN}\equiv  \langle \phi^N_0\phi^N_0  | u^{(ST)}_{NN}(r)  |  \phi^N_0\phi^N_0 \rangle.
\end{align}

For the $\kbar N$ and $\kbar\kbar$ interactions, I consider the $S$-wave interactions of 
the effective single-channel real potentials 
and assume zero-range (delta function) forces 
for simplicity. 
The imaginary part of the $\kbar N$ interaction, which corresponds to the $\pi\Sigma$ decays via the 
$\Sigma$-$\kbar N$ coupling, is omitted.
The $\kbar N$ interaction in the $T=0$ and $T=1$ channels is written as 
\begin{align}\label{eq:vnk}
v^T_{\kbar N}(i',j)&=u^T_{\kbar N}(r_{i'j}) P^{T}_{i'j}
\end{align}
with the delta function $u^T_{\kbar N}(r)=U^T_{\kbar N}\delta(r)$. Here, the isospin-projection operators
can be expressed as 
$P^{T=0}_{kl}=\frac{1-\bvec{\tau}_{k}\cdot\bvec{\tau}_{l}}{4}$ and 
$P^{T=1}_{kl}=\frac{3+\bvec{\tau}_{k}\cdot\bvec{\tau}_{l}}{4}$.
For the $\kbar \kbar$ interaction, 
the spatial-even term exists only in the $T=1$ channel and is given by
\begin{align}\label{eq:vkk}
v^{T=1}_{\kbar \kbar}(i',j')&=u^{T=1}_{\kbar \kbar}(r_{i'j'}) P^{T=1}_{i'j'}
\end{align}
with $u^{T=1}_{\kbar \kbar}(r)=U^{T=1}_{\kbar \kbar}\delta(r)$.
For these zero-range interactions, 
the spatial parts of the expectation values for the
$\kbar N$ and $\kbar\kbar$ pairs in the $0s$-orbit are obtained as 
\begin{align}
&\barV^{T}_{\kbar N}\equiv  \langle \phi^\kbar_0\phi^N_0  | u^{T}_{\kbar N}(r)|\phi^\kbar_0\phi^N_0 \rangle= U^T_{\kbar N}\Bigl(\frac{\lambda}{\pi}\Bigr)^{\frac{3}{2}},
\\
&\barV^{T}_{\kbar\kbar }\equiv  \langle \phi^\kbar_0\phi^\kbar_0  | u^{T}_{\kbar\kbar}(r)  |  \phi^\kbar_0\phi^\kbar_0 \rangle= U^T_{\kbar \kbar} \Bigl(\frac{\nu_K}{\pi}\Bigr)^{\frac{3}{2}}.
\end{align}
The strengths $U^{T=0}_{\kbar N}$, $U^{T=1}_{\kbar N}$, and $U^{T=1}_{\kbar \kbar}$ 
of the interactions are tuned as follows. I first adjust the strength $U^{T=0}_{\kbar N}$ of 
the ${\kbar N}$ interaction in the $T=0$ channel to make the $\Lambda^*$ energy fit with the 
energy $\barT+V^{T=0}_{\kbar N}$ of 
the $\kbar N$ state. 
The strengths $U^{T=1}_{\kbar N}$ and $U^{T=1}_{\kbar \kbar}$ are adjusted to 
reproduce the strength ratios ${\cal F}^{T=1}_{\kbar N}\equiv u^{T=1}_{\kbar N}(r)/u^{T=0}_{\kbar N}(r)$ 
and ${\cal F}^{T=1}_{\kbar \kbar}\equiv u^{T=1}_{\kbar \kbar}(r)/u^{T=0}_{\kbar N}(r)$ 
of the $\kbar N$ and $\kbar\kbar$ 
interactions used in other theoretical works with kaonic nuclei. 
The adopted values of these parameters are explained later. 

\subsection{Parameter settings}

For the $NN$ interaction, I use the 
values of $V_1=-60.65$~MeV,
$V_2=61.14$~MeV, $\eta_1=1.80$~fm, and $\eta_2=1.01$~fm
of the Volkov No.2 parametrization \cite{Volkov:1965zz}. 
I tune the ratio parameters $f^{(ST)}_{NN}$
to fit the experimental data of 
the $S$-wave $NN$-scattering lengths in the spin-triplet and -singlet channels and 
the $\alpha$+$\alpha$-scattering phase shifts and set
values of $f^{(10)}_{NN}=1.3$, $f^{(01)}_{NN}=0.7$, 
$f^{(11)}_{NN}=-0.2$, and $f^{(00)}_{NN}=-0.2$.
This parametrization describes a stronger triplet-even $NN$ interaction to form a bound deuteron state
and a weaker singlet-even $NN$ interaction describing
an unbound $nn$ state. The odd-channel $NN$ interactions are weak repulsions.

In the present calculation, I adopt two sets of parameters of
the $0s$-orbit width $(\nu_N)$ and the strengths of the $\kbar N$ and $\kbar\kbar$ interactions. 
One is the set-I parametrization for the weak-binding case, and 
the other is the set-II parametrization for the deep-binding case. 
In each parametrization, I use fixed $\nu_N$ and $\nu_K$ values 
consistently for all kaonic and normal nuclei.
In the set-I (weak-binding) case, I use $\nu_N=0.16$ fm$^{-2}$, which was
optimized for the deuteron energy in the $0s$-orbital model with the tuned $NN$ interaction. 
In the set-II (deep-binding) case, I choose $\nu_N=0.25$ fm$^{-2}$ 
which reproduces the binding energy and nuclear 
size of the $^4$He system.
To determine the strengths $U^{T=0}_{\kbar N}$, $U^{T=0}_{\kbar \kbar}$, and $U^{T=1}_{\kbar \kbar}$
of the $\kbar N$ and $\kbar\kbar$ interactions, 
I adopt the $\Lambda^*$ energy $(\epsilon_{\Lambda^*})$ and strength ratios ${\cal F}^{T=1}_{\kbar N}$ and 
${\cal F}^{T=1}_{\kbar \kbar}$ that are given by a weak-type chiral interaction 
for the set-I (weak-binding) case, and those that are given by the deep-type phenomenological AY interaction
for the set-II (deep-binding) case.

For the set-I (weak binding) case with $\nu_N=0.16$ fm$^{-2}$, 
I adjust $U^{T=0}_{\kbar N}$ to fit $\epsilon_{\Lambda^*}=-10$~MeV, which 
corresponds to the $\Lambda^*$ resonance-pole position of the chiral SU(3) 
analysis~\cite{Hyodo:2007jq}.
For the strength $U^{T=1}_{\kbar N}$, 
I adopt the value ${\cal F}^{T=1}_{\kbar N}=0.457$ of
the effective single-channel $\kbar N$ potentials
derived from the chiral SU(3) coupled-channel 
analysis. The original ${\kbar N}$ potential in 
Ref.~\cite{Hyodo:2007jq} is energy-dependent and contains imaginary terms, but I omit 
the energy dependence and use only the real part of the interaction
at the $\Lambda^*$ resonance-pole position (1,421~MeV of the $\Lambda^*$ mass). 
For the strength $U^{T=1}_{\kbar \kbar}$, I take the value
${\cal F}^{T=1}_{\kbar \kbar}=-0.345$ of a $\kbar\kbar$ interaction
from Ref.~\cite{Kanada-Enyo:2008wsu}.

For the set-II (deep-binding) case with $\nu_N=0.25$ fm$^{-2}$, 
$U^{T=0}_{\kbar N}$ is adjusted to fit 
$\epsilon_{\Lambda^*}=-27$~MeV 
from the PDG value of $\Lambda^*$ \cite{Tanabashi:2018oca}.
To determine $U^{T=1}_{\kbar N}$ and $U^{T=1}_{\kbar \kbar}$, 
I employ the value ${\cal F}^{T=1}_{\kbar N}=0.294$ of 
the deep-type single-channel AY interaction~\cite{Akaishi:2002bg,Yamazaki:2007cs},
and the value
${\cal F}^{T=1}_{\kbar \kbar}=-0.175$ from Ref.~\cite{Kanada-Enyo:2008wsu}. 

The expectation values of the single-particle kinetic energy and 
the spatial parts of the $NN$, $\kbar N$, and $\kbar\kbar$-interaction terms of the two-particle pairs 
are listed in Table~\ref{tab:energy-count}.
In both the set-I and set-II cases, the ${\kbar N}$ attraction 
is stronger in the $T=0$ channel than in the $T=1$ channel with a factor of 2--3, 
and the $T=1$ ${\kbar \kbar}$ interaction is the weak repulsion. 
Note that the $\kbar N$ and $\kbar\kbar$ interactions adopted here 
are delta forces renormalized to reproduce energy expectation values 
in the present $0s$-orbital model space with a given $\nu_N$ value. 
Such renormalized delta forces cannot be applied to 
variational calculations beyond the assumed model setting.

\begin{table*}[!ht]
\caption{Expectation values and factors for energies of kaonic and normal nuclei in the $0s$-orbital model. 
Upper: the single-particle kinetic energy and 
the spatial part of the expectation values of the interaction terms for the $NN$, $\kbar N$, and $\kbar\kbar$ pairs 
for set-I and set-II parametrization in units of MeV.
Lower: factors of each term as  $A_\textrm{tot}-1$ for the kinetic term, and the product of 
the number of pairs and the isospin component per pair for the interaction terms.
\label{tab:energy-count}}
\begin{center}
\begin{tabular}{cccccccccccc}
\hline
	&	$\barT$	&	$\barV^{(10)}_{NN}$	&	$\barV^{(01)}_{NN}$	&	$\barV^{T=0}_{\kbar N}$	&	$\barV^{T=1}_{\kbar N}$	&	$\barV^{T=0}_{\kbar\kbar}$	\\
set-I  ($\nu_N=0.16$ fm$^{-2}$) &	9.95 	&$	-11.55 	$&$	-6.22 	$&$	-19.95 	$&$	-9.11 	$&	4.59 	\\
set-II  ($\nu_N=0.25$ fm$^{-2}$) &	15.55 	&$	-16.32 	$&$	-8.79 	$&$	-42.55 	$&$	-12.51 	$&	4.97 	\\
&\\
{kaonic nuclei$(J^\pi,T)$}	
 &  $\langle T^\textrm{kin} \rangle$	 & $\langle v^{(10)}_{NN}\rangle$  & $\langle v^{(01)}_{NN}\rangle$  & 
$\langle v^{T=0}_{\kbar N}\rangle$ & $\langle v^{T=1}_{\kbar N}\rangle$  & $\langle v^{T=1}_{\kbar\kbar}\rangle$ \\
$\kbar N (1/2^-,0)$	&	1	&	0	&	0	&	1	&	0	&	0	\\
$\kbar NN (0^-, 1/2)$	&	2	&	0	&	1	&	$2 (\frac{3}{4})$	&	$2 (\frac{1}{4})$	&	0	\\
$\kbar NN (0^-, 3/2)$	&	2	&	0	&	1	&	0	&	2	&	0	\\
$\kbar NN (1^-, 1/2)$	&	2	&	1	&	0	&	$2(\frac{1}{4})$	&	$2 (\frac{3}{4})$	&	0	\\
$\kbar\kbar N (1/2^+, 1/2)$	&	2	&	0	&	0	&	$2 (\frac{3}{4})$	&	$2 (\frac{1}{4})$	&	1	\\
$\kbar\kbar NN (0^+,0)$	&	3	&	0	&	1	&	$4 (\frac{3}{4})$	&	$4 (\frac{1}{4})$	&	1	\\
$\kbar\kbar NN (0^+,1)$	&	3	&	0	&	1	&	$4 (\frac{1}{2})$	&	$4 (\frac{1}{2})$	&	1	\\
$\kbar\kbar NN (0^+,2)$	&	3	&	0	&	1	&	0	&	4	&	1	\\
$\kbar\kbar NN(1^+,1)$	&	3	&	1	&	0	&	$4 (\frac{1}{4})$	&	$4 (\frac{3}{4})$	&	1	\\
$\kbar NNN (1/2^-,0)$	&	3	&	$3(\frac{1}{2})$	&	$3(\frac{1}{2})$	&	$3 (\frac{1}{2})$	&	$3 (\frac{1}{2})$	&	0	\\
$\kbar NNNN (0^-,1/2)$	&	4	&	$6(\frac{1}{2})$	&	$6(\frac{1}{2})$	&	$4 (\frac{1}{4})$	&	$4 (\frac{3}{4})$	&	0	\\
&\\
{nuclei$(J^\pi,T)$}		&	 $\langle T^\textrm{kin} \rangle$	 & $\langle v^{(10)}_{NN}\rangle$  & $\langle v^{(01)}_{NN}\rangle$ &		&		&		\\
$NN (0^+,1)$	&	1	&	0	&	1	&		&		&		\\
$NN (1^+,0)$	&	1	&	1	&	0	&		&		&		\\
$NNN (1/2^+,1/2)$	&	2	&	$3(\frac{1}{2})$	&	$3(\frac{1}{2})$	&		&		&		\\
$NNNN (0^+,0)$	&	3	&	$6(\frac{1}{2})$	&	$6(\frac{1}{2})$	&		&		&		\\
\hline
\end{tabular}
\end{center}
\end{table*}

\section{Results of kaonic nuclei with the $0s$-orbital model} \label{sec:results-1}

\subsection{Energy counting}
With the present $0s$-orbital model, 
I calculate the energies $E_{\kbar N^A}^{(J^\pi,T)}$ and $E_{\kbar\kbar N^A}^{(J^\pi,T)}$ 
of the $(J^\pi.T)$ states of the kaonic nuclei, $\kbar N^A$ and $\kbar \kbar N^A$.
These are obtained by counting the spin and isospin components of the $NN$, $\kbar N$, and $\kbar\kbar$ pairs
and can be expressed  simply with the expectation value terms, $\barT$, ${\barV}^{(ST)}_{NN}$, ${\barV}^{T}_{\kbar N}$, 
and ${\barV}^{T}_{\kbar \kbar}$. Hence, I obtain
the lowest $(J^\pi,T)$ states of each system of 
the $\kbar N(1/2^-,0)$, $\kbar NN(0^-,1/2)$,
$\kbar\kbar N(1/2^+,1/2)$, $\kbar\kbar NN(0^+,0)$, $\kbar NNN(1/2^-,0)$, and $\kbar NNNN(0^-,1/2)$. 

The energy of $\kbar N(1/2^-,0)$ corresponding to the $\Lambda^*$ state is 
\begin{align}
E_{\kbar N}^{(1/2^-,0)}=\barT+{\barV}^{T=0}_{\kbar N}= \epsilon_{\Lambda^*},
\end{align}
which is used as an input to determine the interaction strengths in the present framework.
For $\kbar NN(0^-,1/2)$,
$\kbar\kbar N(1/2^+,1/2)$, $\kbar\kbar NN(0^+,0)$, the energies are given by
\begin{align}
\label{eq:E-NNK}
&E_{\kbar NN}^{(0^-,1/2)}=
2\barT+\barV^{(01)}_{NN}+2\Bigl(\frac{3}{4}\barV^{T=0}_{\kbar N}\Bigr)+2\Bigl(\frac{1}{4}\barV^{T=1}_{\kbar N}\Bigr)\nonumber\\
&\qquad\qquad =\epsilon_{\Lambda^*}+\epsilon_{nn}+\frac{1}{2}\barV^{T=0}_{\kbar N}+\frac{1}{2}\barV^{T=1}_{\kbar N},
\\
\label{eq:E-NKK}
&E^{(1/2^+,1/2)}_{\kbar\kbar N}=2\barT+2\Bigl(\frac{3}{4}\barV^{T=0}_{\kbar N}\Bigr)+2\Bigl(\frac{1}{4}\barV^{T=1}_{\kbar N}\Bigr)+\barV^{T=1}_{\kbar\kbar}\nonumber\\
&\qquad=\epsilon_{\Lambda^*}+\barT+\frac{1}{2}\barV^{T=0}_{\kbar N}+\frac{1}{2}\barV^{T=1}_{\kbar N}
+\barV^{T=1}_{\kbar\kbar},\\
\label{eq:E-NNKK}
&E^{(0^+,0)}_{\kbar\kbar NN}=3\barT+\barV^{(01)}_{NN}\nonumber\\
&\qquad\qquad\quad +4\Bigl(\frac{3}{4}\barV^{T=0}_{\kbar N}\Bigr)+4\Bigl(\frac{1}{4}\barV^{T=1}_{\kbar N}\Bigr)+\barV^{T=1}_{\kbar\kbar}\nonumber\\
&\qquad\qquad =2\epsilon_{\Lambda^*}+\epsilon_{nn}+\barV^{T=0}_{\kbar N}+\barV^{T=1}_{\kbar N}
+\barV^{T=1}_{\kbar\kbar},
\end{align}
where  $\epsilon_{nn}= \barT+\barV^{(01)}_{NN}$ is the energy of a two-neutron state with 
the $0s$-orbital  configuration and has a positive value.
The energies of $\kbar NNN(1/2^-,0)$ and $\kbar NNNN(0^-,1/2)$ are written as 
\begin{align}
\label{eq:E-NNNK}
&E_{\kbar NNN}^{(1/2^-,0)}=3\barT+3\Bigl(\frac{1}{2}\barV^{(01)}_{NN}\Bigr)+3\Bigl(\frac{1}{2}\barV^{(01)}_{NN}\Bigr) \nonumber\\
&\qquad\qquad\quad
+3\Bigl(\frac{1}{2}\barV^{T=0}_{\kbar N}\Bigr)
+3\Bigl(\frac{1}{2}\barV^{T=1}_{\kbar N}\Bigr)\nonumber\\
&\qquad\qquad =\epsilon_{\Lambda^*}+\epsilon_{^3\textrm{H}}+\frac{1}{2}\barV^{T=0}_{\kbar N}+\frac{3}{2}\barV^{T=1}_{\kbar N},\\
\label{eq:E-NNNNK}
&E_{\kbar NNNN}^{(0^-,1/2)}=4\barT+6\Bigl(\frac{1}{2}\barV^{(01)}_{NN}\Bigr)+6\Bigl(\frac{1}{2}\barV^{(01)}_{NN}\Bigr) \nonumber\\
&\qquad\qquad\quad
+4\Bigl(\frac{1}{4}\barV^{T=0}_{\kbar N}\Bigr)
+4\Bigl(\frac{3}{4}\barV^{T=1}_{\kbar N}\Bigr)\nonumber\\
&\qquad\qquad =\epsilon_{\Lambda^*}+\epsilon_{^4\textrm{He}}+3\barV^{T=1}_{\kbar N},
\end{align}
where 
$\epsilon_{^3\textrm{H}}=2\barT+\frac{3}{2}\barV^{(10)}_{NN}+\frac{3}{2}\barV^{(01)}_{NN}$ and 
$\epsilon_{^4\textrm{He}}=3\barT+3\barV^{(10)}_{NN}+3\barV^{(01)}_{NN}$
are the energies of the $^3\textrm{H}$ and 
$^4\textrm{He}$ nuclei 
with the $0s$-orbital  configuration, respectively.

The energy counting
for the lowest and other $(J^\pi,T)$ states is summarized 
in Table~\ref{tab:energy-count}. The factor for  each interaction term is given by 
the product of the number of pairs and the spin-isospin component per pair.
The strong $\kbar N$ interaction in the $T=0$ channel generally 
induces isoscalar $\kbar N$ correlation. 
On the other hand, for $NN$ pairs, an isoscalar $NN$ correlation is favored because 
the triplet-even $(ST)=(10)$ term is stronger than the singlet-even $(ST)=(01)$ term 
in the effective $NN$ interaction.
Furthermore, nuclear systems in the $0s$-orbit favor spin and/or isospin saturation
as in the $^4$He system  
because of the Pauli principle of nucleons.
In kaonic nuclei, 
the isoscalar $\kbar N$ and $NN$ correlations compete against each other.
In $A=2$ kaonic nuclei, 
the $\kbar NN(0^-,1/2)$ and $\kbar\kbar NN(0^+,0)$ states containing 
an isovector $(ST)=(01)$ $NN$ pair are energetically favored 
over the $\kbar NN(1^-,1/2)$ and $\kbar\kbar NN(1^+,1)$ states with a dueteron-like 
$(ST)=(10)$ $NN$ pair, indicating that 
isoscalar $\kbar N$ correlation is superior to isoscalar $(ST)=(10)$ $NN$ correlation.
In kaonic nuclei with  $A\ge 3$, the isospin saturation occurs in the nuclear part; consequently,
the isoscalar $\kbar N$ correlation gradually decreases with the increase of $A$, as can be seen in the 
reduction of the $T=0$ component of the $\kbar N$ pairs. 
The fraction of the $T=0$ component is 1 in the $\kbar N(1/2^-,0)$ state, $\frac{3}{4}$ in the $\kbar NN(0^-,1/2)$, $\kbar \kbar N(1/2^+,1/2)$, 
and $\kbar\kbar NN(0^+,0)$ states, $\frac{1}{2}$ in the  $\kbar NNN(1/2^-,0)$ state, and 
$\frac{1}{4}$ in the $\kbar NNNN(0^-,1/2)$ state.

\subsection{Energy spectra of kaonic nuclei}

The calculated energies 
obtained using set-I (weak-binding) and set-II (deep-binding) are listed in Tables~\ref{tab:ene-cal1}
and \ref{tab:ene-cal2}, respectively.
For kaonic nuclei, the total energies $(E)$, $\kbar$-separation energies $({\cal S}_\kbar)$, and 
$\Lambda^*$-separation energies $({\cal S}_{\Lambda^*})$ are shown. 
For normal nuclei, the total energies, nucleon-separation energies $({\cal S}_N)$, and 
deuteron-separation energies $({\cal S}_{d})$ are shown. 
Moreover, the contributions of the kinetic energy and $NN$, $\kbar N$, and $\kbar\kbar$-interaction terms 
are listed in the table. 

\begin{table*}[!ht]
\caption{The energies of the $(J^\pi,T)$ states of kaonic and normal nuclei calculated by
the $0s$-orbital model with the set-I (weak-binding) parametrization.
The total energy $(E=-\textrm{B.E.})$ and the contributions of kinetic ($ T^\textrm{kin}$), 
$NN$ ($v_{NN} $), $\kbar N$  ($  v_{\kbar N} $), and 
$\kbar\kbar$  ($  v_{NN}$) interactions are listed. 
Separation energies ${\cal S}_{\kbar}$ and ${\cal S}_{\Lambda^*}$ for kaonic nuclei 
and  ${\cal S}_{N}$ and ${\cal S}_{d}$ for normal nuclei are also shown. Energies are in units of MeV.
 \label{tab:ene-cal1}}
\begin{center}
\begin{tabular}{c|cccc|cccc}
\hline
\multicolumn{4}{c}{set-I ($\epsilon_{\Lambda^*}=-10$~MeV, $\nu_N=0.16$~fm$^{-2}$)}\\
	&	$\langle T^\textrm{kin}\rangle$	&	$\langle v_{NN}\rangle$	&	$\langle v_{\kbar N} \rangle$	&	$\langle v_{\kbar\kbar}\rangle $	&	$E$	&	${\cal S}_{\kbar}$	&	${\cal S}_{\Lambda^*}$	\\
$\kbar N (1/2^-,0)$	&	10.0 	&$	0.0 	$&$	-20.0 	$&$	0.0 	$&$	-10.0 	$&$	10.0 	$&$	-	$\\
$\kbar NN (0^-, 1/2)$	&	19.9 	&$	-6.2 	$&$	-34.5 	$&$	0.0 	$&$	-20.8 	$&$	{(20.8)}	$&$	10.8 	$\\
$\kbar NN (0^-, 3/2)$	&	19.9 	&$	-6.2 	$&$	-18.2 	$&$	0.0 	$&$	-4.5 	$&$	{(4.5)}	$&$	-	$\\
$\kbar NN (1^-, 1/2)$	&	19.9 	&$	-11.6 	$&$	-23.6 	$&$	0.0 	$&$	-15.3 	$&$	13.7 	$&$	5.3 	$\\
$\kbar\kbar N (1/2^+, 1/2)$	&	19.9 	&$	0.0 	$&$	-34.5 	$&$	4.6 	$&$	-10.0 	$&$	-0.01 	$&$	-0.01 	$\\
$\kbar\kbar NN (0^+,0)$	&	29.9 	&$	-6.2 	$&$	-69.0 	$&$	4.6 	$&$	-40.7 	$&$	19.9 	$&$	20.7 	$\\
$\kbar\kbar NN (0^+,1)$	&	29.9 	&$	-6.2 	$&$	-58.1 	$&$	4.6 	$&$	-29.9 	$&$	9.1 	$&$	-	$\\
$\kbar\kbar NN (0^+,2)$	&	29.9 	&$	-6.2 	$&$	-36.5 	$&$	4.6 	$&$	-8.2 	$&$	3.7 	$&$	-	$\\
$\kbar\kbar NN(1^+,1)$	&	29.9 	&$	-11.6 	$&$	-47.3 	$&$	4.6 	$&$	-24.4 	$&$	9.1 	$&$	-	$\\
$\kbar NNN (1/2^-,0)$	&	29.9 	&$	-26.7 	$&$	-43.6 	$&$	0.0 	$&$	-40.4 	$&$	33.6 	$&$	38.8 	$\\
$\kbar NNNN (0^-,1/2)$	&	39.8 	&$	-53.3 	$&$	-47.3 	$&$	0.0 	$&$	-60.8 	$&$	37.3 	$&$	54.0 	$\\
\multicolumn{5}{c}{}\\															
	&	$ \langle T^\textrm{kin}\rangle$	&	$ \langle v_{NN}\rangle$	&		&		&	$E$	&	${\cal S}_{N}$	&	${\cal S}_{d}$	\\
$NN (0^+,1)$	&	10.0 	&$	-6.2 	$&$		$&$		$&$	3.7 	$&$	-	$&$	-	$\\
$NN (1^+,0)$	&	10.0 	&$	-11.6 	$&$		$&$		$&$	-1.6 	$&$	1.6 	$&$	-	$\\
$NNN (1/2^+,1/2)$	&	19.9 	&$	-26.7 	$&$		$&$		$&$	-6.7 	$&$	5.2 	$&$	-	$\\
$NNNN (0^+,0)$	&	29.9 	&$	-53.3 	$&$		$&$		$&$	-23.5 	$&$	16.7 	$&$	20.3 	$\\
\hline
\end{tabular}
\end{center}
\end{table*}

\begin{table*}[!ht]
\caption{Same as Table~\ref{tab:ene-cal1}, but results are calculated with 
the set-II (deep-binding) parametrization. 
 \label{tab:ene-cal2}}
\begin{center}
\begin{tabular}{c|cccc|cccc}
\hline
\multicolumn{4}{c}{set-II ($\epsilon_{\Lambda^*}=-27$~MeV, $\nu_N=0.25$~fm$^{-2}$)}\\
	&	$T^\textrm{kin}$	&	$v_{NN}$	&	$ v_{\kbar N} $	&	$v_{\kbar\kbar} $	&	$E$	&	${\cal S}_{\kbar}$	&	${\cal S}_{\Lambda^*}$	\\
$\kbar N (1/2^-,0)$	&	15.6 	&$	0.0 	$&$	-42.6 	$&$	0.0 	$&$	-27.0 	$&$	27.0 	$&$	-	$\\
$\kbar NN (0^-, 1/2)$	&	31.1 	&$	-8.8 	$&$	-70.1 	$&$	0.0 	$&$	-47.8 	$&$	{(47.8)}$&$	20.8 	$\\
$\kbar NN (0^-, 3/2)$	&	31.1 	&$	-8.8 	$&$	-25.0 	$&$	0.0 	$&$	-2.7 	$&$	{(2.7)}$&$	-	$\\
$\kbar NN (1^-, 1/2)$	&	31.1 	&$	-16.3 	$&$	-40.0 	$&$	0.0 	$&$	-25.3 	$&$	24.5 	$&$	-1.7 	$\\
$\kbar\kbar N (1/2^+, 1/2)$	&	31.1 	&$	0.0 	$&$	-70.1 	$&$	5.0 	$&$	-34.0 	$&$	7.0 	$&$	7.0 	$\\
$\kbar\kbar NN (0^+,0)$	&	46.7 	&$	-8.8 	$&$	-140.2 	$&$	5.0 	$&$	-97.3 	$&$	49.6 	$&$	43.3 	$\\
$\kbar\kbar NN (0^+,1)$	&	46.7 	&$	-8.8 	$&$	-110.1 	$&$	5.0 	$&$	-67.3 	$&$	19.5 	$&$	-	$\\
$\kbar\kbar NN (0^+,2)$	&	46.7 	&$	-8.8 	$&$	-50.1 	$&$	5.0 	$&$	-7.2 	$&$	4.5 	$&$	-	$\\
$\kbar\kbar NN(1^+,1)$	&	46.7 	&$	-16.3 	$&$	-80.1 	$&$	5.0 	$&$	-44.8 	$&$	19.5 	$&$	-	$\\
$\kbar NNN (1/2^-,0)$	&	46.7 	&$	-37.7 	$&$	-82.6 	$&$	0.0 	$&$	-73.6 	$&$	67.0 	$&$	72.8 	$\\
$\kbar NNNN (0^-,1/2)$	&	62.2 	&$	-75.3 	$&$	-80.1 	$&$	0.0 	$&$	-93.2 	$&$	64.5 	$&$	86.7 	$\\
\multicolumn{5}{c}{}\\															
	&	$T^\textrm{kin}$	&	$ v_{NN} $	&		&		&	$E$	&	${\cal S}_{N}$	&	${\cal S}_{d}$	\\
$NN (0^+,1)$	&	15.6 	&$	-8.8 	$&$		$&$		$&$	6.8 	$&$	-	$&$	-	$\\
$NN (1^+,0)$	&	15.6 	&$	-16.3 	$&$		$&$		$&$	-0.8 	$&$	0.8 	$&$	-	$\\
$NNN (1/2^+,1/2)$	&	31.1 	&$	-37.7 	$&$		$&$		$&$	-6.6 	$&$	5.8 	$&$	-	$\\
$NNNN (0^+,0)$	&	46.7 	&$	-75.3 	$&$		$&$		$&$	-28.7 	$&$	22.1 	$&$	27.1 	$\\
\hline
\end{tabular}
\end{center}
\end{table*}

In the lowest states, i.e., 
 $\kbar N(1/2^-,0)$, $\kbar NN(0^-,1/2)$,
  $\kbar NNN(1/2^-,0)$, $\kbar NNNN(0^-,1/2)$, and $\kbar\kbar NN(0^+,0)$, 
an antikaon and a $\Lambda^*$ are deeply bound 
due to the remarkable contribution of the $\kbar N$ interaction 
with ${\cal S}_{\Lambda^*}\gtrsim 10$~MeV in the set-I result 
and ${\cal S}_{\Lambda^*}\gtrsim 20$~MeV in the set-II result.
An exception is the  $\kbar \kbar N(1/2^+,1/2)$, which is 
almost bound  close to the 
$\kbar+\Lambda^*$-threshold energy in the set-I result and is weakly 
bound with ${\cal S}_{\Lambda^*}=7.0$~MeV
in the set-II result.
The other $(J^\pi,T)$ states of the kaonic nuclei are relatively unfavored because they 
have weaker isoscalar $\kbar N$ correlations than the lowest states.

I compare the present results with other theoretical results 
in Table~\ref{tab:NNKK}.
The binding energies of the lowest states are compared with the 
theoretical values from Refs.~\cite{Barnea:2012qa,Maeda:2013zha,Ohnishi:2017uni}
of the few-body calculations using 
weak-type and deep-type $\kbar N$ interactions. 
The energy spectra of the set-I result agrees reasonably well with the   
results of other calculations with weak-type chiral interactions, 
and the set-II result corresponds well with other theoretical results with deep-type AY interactions. 
In Table~\ref{tab:NNKK}, I also compare 
the present results for rms distances $R_{NN}$,  $R_{\kbar N}$, and  $R_{\kbar\kbar}$ 
for $NN$, $\kbar N$, and $\kbar\kbar$ pairs in kaonic nuclei with other theoretical results. 
The rms distances are constant for a fixed $\nu_N$ value in the present $0s$-orbital model, whereas
they are dependent on the system in other calculations with few-body approaches that include dynamical effects. 
Nevertheless, the present calculations using set-I and set-II yield reasonable results for 
$R_{NN}$,  $R_{\kbar N}$, and  $R_{\kbar\kbar}$ in kaonic nuclei that are comparable to other calculations
of weak-type and deep-type interactions, respectively.  
Hence, the present choices of $\nu_N$ adopted for sets-I and II
are reasonable for global descriptions of the 
system sizes of kaonic nuclei.

\begin{figure*}[!htp]
\includegraphics[width=12 cm]{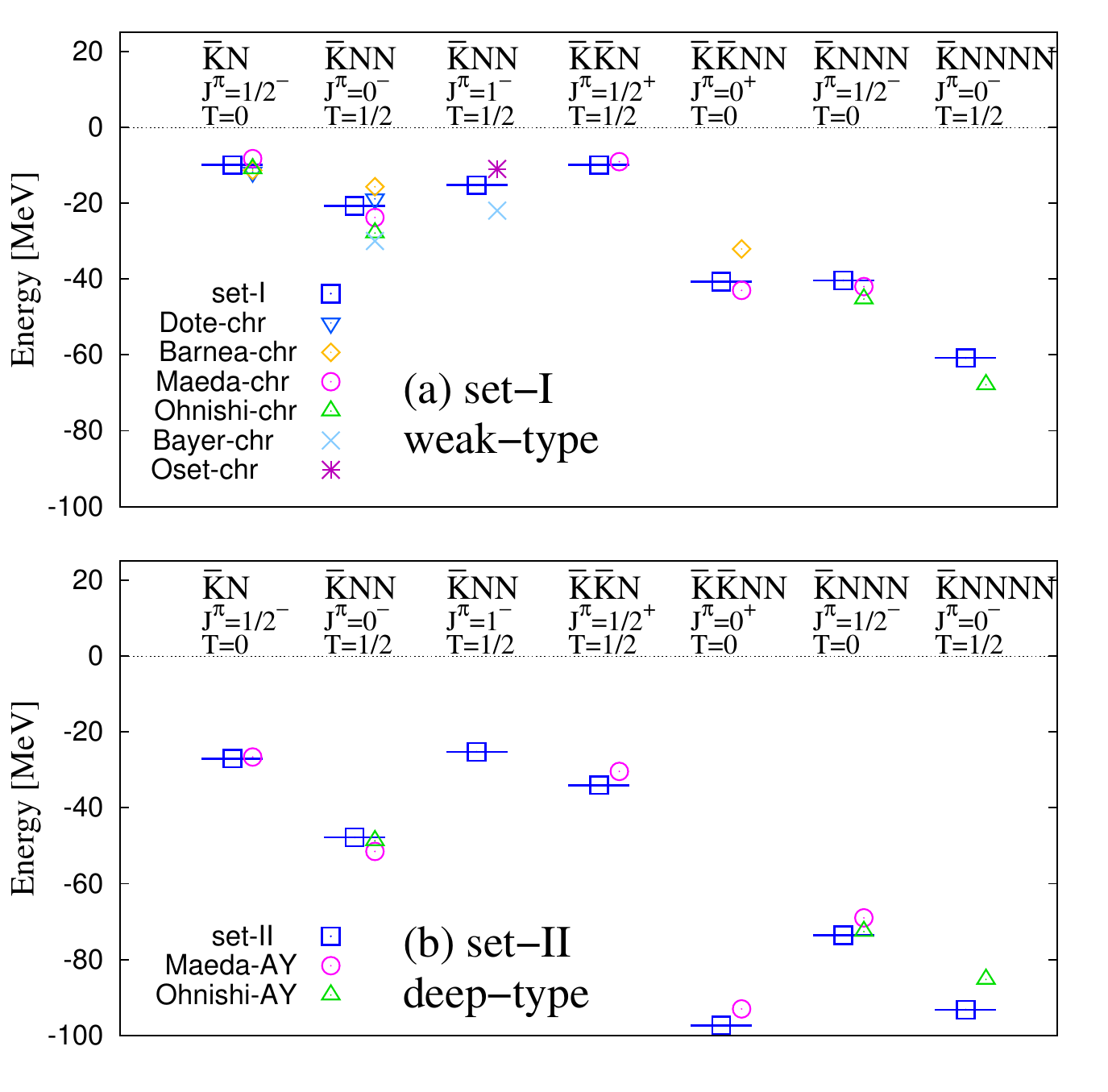}
\caption{Energies $E=-\textrm{B.E.}$ of the kaonic nuclei. 
(a) Upper: the set-I (weak-binding) results are shown together with
other theoretical results obtained by few-body approaches using weak-type chiral 
interactions by  Dot\'e {\it et al.} (ORB type-I case)~\cite{Dote:2008hw},
Barnea {\it et al.} (BGL case)~\cite{Barnea:2012qa}, Maeda {\it et al.} 
(weak-chiral-regime case)~\cite{Maeda:2013zha}, 
Ohnishi {\it et al.} (Kyoto type-I case)~\cite{Ohnishi:2017uni}, 
Bayar {\it et al.} (normal-radius case)~\cite{Bayar:2011qj}, 
and Oset {\it et al.} (normal-radius case)~\cite{Oset:2012gi}. 
(b) Lower: the set -II (deep-binding) results are shown together with
other theoretical results for the deep-type AY interaction 
by Maeda {\it et al.}~\cite{Maeda:2013zha} and 
Ohnishi {\it et al.}~\cite{Ohnishi:2017uni}.
	\label{fig:knuclei-spe}}
\end{figure*}

In Fig.~\ref{fig:knuclei-spe},
the energy spectra of kaonic nuclei 
are shown together with other theoretical results. 
Figure~\ref{fig:knuclei-spe}(a) shows the set-I (weak-binding) result in comparison with 
other theoretical results for weak-type chiral interactions from
Refs.~\cite{Dote:2008hw,Barnea:2012qa,Maeda:2013zha,Ohnishi:2017uni,Bayar:2011qj,Oset:2012gi}. 
Figure~\ref{fig:knuclei-spe}(b) shows the set-II (deep-binding) result
compared with other calculations for the deep-type AY interaction. 
In each group of weak- and deep-type calculations, the present calculation 
describes the energy systematics of other theoretical results.
This means that the binding energies of kaonic nuclei are not very sensitive to the details of the $\kbar N$ interaction but essentially depend upon the energy of the $\kbar N$ bound state corresponding to $\Lambda^*$.
Moreover, 
the leading part of the binding energies 
may be understood by simple energy counting in the present $0s$-orbital model, in which 
the spin and isospin symmetries play essential roles  
in the binding mechanism of light-mass kaonic nuclei.

\begin{table*}[!ht]
\caption{Binding energies (B.E.) and rms distances of  $NN$~($R_{N N}$), 
$\kbar N$~($R_{\kbar N}$), and $\kbar\kbar$~($R_{\kbar \kbar}$) pairs in the lowest states of kaonic
and normal nuclei. 
Calculated values obtained 
with the set-I (weak-binding) and set-II (deep-binding) cases are 
compared with other theoretical results 
obtained with weak-type chiral and deep-type AY interactions 
by Maeda {\it et al.}~\cite{Maeda:2013zha}, 
Ohnishi {\it et al.}~\cite{Ohnishi:2017uni}, and Barnea {\it et al.}~\cite{Barnea:2012qa}.
For values of weak-type chiral interactions, 
the weak-chiral-regime result of Ref.~\cite{Maeda:2013zha}, 
the Kyoto type-I result of Ref.~\cite{Ohnishi:2017uni}, and 
the BGL result of Ref.~\cite{Barnea:2012qa} are listed.
For normal nuclei, the AV4' result obtained by Ohnishi {\it et al.} 
from Ref.~\cite{Ohnishi:2017uni} is also shown.
\label{tab:NNKK}}
\begin{center}
\begin{tabular}{ccrrrrrrrr}
\hline
		&		&\multicolumn{2}{c}{present}&\multicolumn{2}{c}{Maeda~\cite{Maeda:2013zha}}	&\multicolumn{2}{c}{Ohnishi~\cite{Ohnishi:2017uni}}	&
		Barnea~\cite{Barnea:2012qa}\\
	&		&	set-I	&	set-II	&	Chiral	&	AY	&	Chiral	&	AY	&	Chiral	\\
$\nu_N$ (fm$^{-2}$) &		&	0.16	&	0.25	&		&		&		&		&		\\
\multicolumn{2}{l}{kaonic nuclei$(J^\pi,T)$}	&\\															
$\kbar N (1/2^-,0)$	&	B.E.~(MeV)	&	10	&	27	&	8.3	&	26.6	&		&		&	11.4	\\
	&	$R_{\kbar N}$~(fm)	&	2.17 	&	1.73 	&	2.25	&	1.41	&		&		&	1.87	\\
	&\\																
$\kbar NN (0^-, 1/2)$	&	B.E.~(MeV)	&	20.8 	&	47.8 	&	23.8	&	51.5	&	27.9	&	48.7	&	15.7	\\
	&	$R_{NN}$~(fm)	&	1.77 	&	1.41 	&	1.93	&	1.62	&	2.16	&	1.84	&		\\
	&	$R_{\kbar N}$~(fm)	&	2.17 	&	1.73 	&		&		&	1.80	&	1.55	&		\\
	&\\																
$\kbar\kbar NN; (0^+,0)$	&	B.E.~(MeV)	&	40.7 	&	97.3 	&	43	&	93	&		&		&	32.1	\\
	&	$R_{NN}$~(fm)	&	1.77 	&	1.41 	&	1.57	&	1.35	&		&		&	1.84	\\
	&	$R_{\kbar\kbar}$~(fm)	&	2.50 	&	2.00 	&		&		&		&		&	2.31	\\
	&\\																
$\kbar NNN (1/2^-,0)$	&	B.E.~(MeV)	&	40.4 	&	73.6 	&	42	&	69	&	45.3	&	72.6	&		\\
	&	$R_{NN}$~(fm)	&	1.77 	&	1.41 	&	1.89	&	1.75	&	1.99	&	1.87	&		\\
	&	$R_{\kbar N}$~(fm)	&	2.17 	&	1.73 	&		&		&	1.79	&	1.63	&		\\
	&\\																
$\kbar NNNN (0^-,1/2)$	&	B.E.~(MeV)	&	60.8 	&	93.2 	&		&		&	67.9	&	85.2	&		\\
	&	$R_{NN}$~(fm)	&	1.77 	&	1.41 	&		&		&	1.98	&	2.07	&		\\
	&	$R_{\kbar N}$~(fm)	&	2.17 	&	1.73 	&		&		&	1.83	&	1.81	&		\\
\multicolumn{2}{l}{nuclei$(J^\pi,T)$}	&\\																
$NN (1^+,0)$	&	B.E.~(MeV)	&	1.6 	&	0.8 	&		&		&	2.24	&	2.24	&		\\
	&	$R_{NN}$~(fm)	&	1.77 	&	1.41 	&		&		&	4.04	&	4.04	&		\\
$NNN (1/2^+,1/2)$	&	B.E.~(MeV)	&	6.7 	&	6.6 	&		&		&	8.99	&	8.99	&		\\
$NNNN (0^+,0)$	&	B.E.~(MeV)	&	23.5 	&	28.7 	&		&		&	32.1	&	32.1	&		\\
\hline
\end{tabular}
\end{center}
\end{table*}

\subsection{$\kbar NN$ system}
In the $\kbar NN$ system, 
the $(J^\pi,T)=(0^-,1/2)$ state is the lowest and has been investigated by 
many groups as a deeply bound $K^- pp$ system. For this state, 
I obtain a binding energy that approximately two times larger 
than the $\Lambda^*$-binding energy (see Table~\ref{tab:ene-cal1} and Fig.~\ref{fig:knuclei-spe}(a) for 
the set-I result, and Table~\ref{tab:ene-cal2} and Fig.~\ref{fig:knuclei-spe}(b) for the set-II result).
This energy relation 
$E_{\kbar NN}^{(0^-,1/2)}\approx 2\epsilon_{\Lambda^*}$ is naively understood
by the energies for two $\kbar N$ pairs in the ${\kbar NN}(0^-,1/2)$ state. Quantitatively, 
it can be described by the present model with energy counting as 
\begin{align}
&E_{\kbar NN}^{(0^-,1/2)}=
2\epsilon_{\Lambda^*}+\frac{1}{2}(\barV^{T=1}_{\kbar N}-\barV^{T=0}_{\kbar N})+\barV^{(01)}_{NN}, 
\end{align}
meaning that the $NN$ attraction compensates for the energy loss by reducing the isoscalar
$\kbar N$ correlation in $\kbar NN(0^-,1/2)$. 
It is interesting to compare the energies of ${\kbar NN}(0^-,1/2)$ and ${\kbar\kbar N}{(1/2^+,1/2)}$. The
former is deeply bound and the latter is weakly (or almost) bound, 
even though the two systems have the same degree of $\kbar N$ interaction.
As shown in Eqs.~\eqref{eq:E-NNK} and \eqref{eq:E-NKK}, 
the energy difference is just the last term, $\barV^{(01)}_{NN}$ of the $NN$ attraction 
in ${\kbar NN}(0^-,1/2)$ and 
the $\barV^{T=1}_{\kbar\kbar}$ term of the $\kbar\kbar$ repulsion 
in ${\kbar\kbar N}{(1/2^+,1/2)}$. This brings about a significant difference 
in the $\Lambda^*$-separation energies of two systems indicating 
the important role of the singlet-even $NN$ 
attraction in the binding mechanism of the $\kbar NN(0^-,1/2)$ state.

The $\kbar NN(1^-,1/2)$ state with a deuteron-like $(ST)=(10)$ $NN$ pair 
has a higher energy than the lowest $\kbar NN(0^-,1/2)$ state having a $(ST)=(01)$ $NN$ pair, 
despite the triplet-even $NN$ interaction being stronger than the 
singlet-even $NN$ interaction. This is   
because  there is no isoscalar $\kbar N$ correlation 
in $\kbar NN(1^-,1/2)$ with a fraction $1/4$ of the $T=0$ component.
In the set-I result, the $\kbar NN(1^-,1/2)$ state is 
weakly bound with $S_{\Lambda^*}=5.3$~MeV. This result is consistent with 
the prediction of $S_{\Lambda^*}=9$~MeV for a coupled-channel Faddeev calculation~\cite{Oset:2012gi}.
In the set-II result, I obtained $E_{\kbar NN}^{(1^-,1/2)}=-25.3$~MeV, which is slightly higher than 
the $\Lambda^*$-decay threshold at $-27.0$~MeV. 

\subsection{$\kbar\kbar NN$ system}

The $(J^\pi,T)=(0^+,0)$ state is the lowest of the $\kbar\kbar NN$ system, 
and is deeply
bound because of the strong isoscalar $\kbar N$ correlation with a fraction of 3/4 for the $T=0$ component.
In both the set-I and II results, 
the energy of $\kbar\kbar NN(0^+,0)$ is approximately 
twice that of ${\kbar NN}{(0^-,1/2)}$, meaning that the $\kbar$-separation energy 
is almost constant between the two systems.
This energy relation,
\begin{align}\label{eq:E-NNKK-NNK}
&E_{\kbar\kbar NN}^{(0^+,0)}\approx 2E_{\kbar NN}^{(0^-,1/2)}, 
\end{align}
is also roughly satisfied in other theoretical results for Refs.~\cite{Barnea:2012qa,Maeda:2013zha}. 
In the present $0s$-orbital model, it is easy to see the energy relation from Eqs.~\eqref{eq:E-NNK}
and \eqref{eq:E-NNKK} as
\begin{align}
&E^{(0^+,0)}_{\kbar\kbar NN}=2E_{\kbar NN}^{(0^-,1/2)}-\frac{1}{2}\epsilon_{nn}+\barV^{T=1}_{\kbar\kbar},
\end{align}
where the last two terms yield minor contributions 
as $-\frac{1}{2}\epsilon_{nn}=-1.9$~MeV and $\barV^{T=1}_{\kbar\kbar}=4.6$~MeV for the set-I case
($-\frac{1}{2}\epsilon_{nn}=-3.4$~MeV and $\barV^{T=1}_{\kbar\kbar}=5.0$~MeV for the set-II case) 
and cancel each other.

Let me discuss the binding mechanism of a singlet-even $NN$ pair in the $\kbar NN$ and $\kbar\kbar NN$ systems
in a Born-Oppenheimer picture of light-mass antikaons around heavy-mass nucleons. 
Two nucleons in the singlet-even channel are unbound without antikaons, 
but they are deeply bound by a surrounding antikaon 
in the $\kbar NN$ system and further 
deeply bound by two antikaons in the $\kbar\kbar NN$ system.
The mechanism for binding the two nucleons by an antikaon in the $\kbar^- pp$ system
was originally interpreted as a super-strong nuclear force 
caused by a  migrating $\kbar$ meson by Yamazaki and Akaishi~\cite{Yamazaki:2007cs,Yamazaki:2007hj}. 
In the perturbative picture, the constant ${\cal S}_{\kbar}$ in the  $\kbar NN$ and  $\kbar\kbar NN$ systems
can be described by the condensation of two antikaons in the same orbit around two nucleons. 
if the $\kbar\kbar$ interaction is minor. 
It should be noted that, when three antikaons around two nucleons are considered, 
the additional(third) antikaon no longer exhibits isoscalar $\kbar N$ correlation 
because the isospin is already saturated in the $\kbar\kbar NN$ system. 

The higher states $\kbar\kbar NN(0^+,1)$, $\kbar\kbar NN(0^+,2)$, and $\kbar\kbar NN(1^+,1)$, 
exhibit weaker isoscalar $\kbar N$ correlations
because these state have lower symmetry in the isospin coupling
between antikaons and nucleons than $\kbar\kbar NN(0^+,0)$  does. In particular, 
the $\kbar\kbar NN(0^+,1)$ and $\kbar\kbar NN(0^+,2)$ states 
are composed of isovector $NN$ and $\kbar N$ pairs coupled to $T=1$ and $T=2$, respectively, 
and the $\kbar\kbar NN(1^+,1)$ state contains an isoscalar $NN$ pair.

Comparing $\kbar\kbar NN(0^+,1)$ and $\kbar\kbar NN(1^+,1)$, 
one can see the competition between the isoscalar $\kbar N$ and $NN$ correlations. 
$\kbar\kbar NN(0^+,1)$ has a moderate isoscalar $\kbar N$ component with a fraction of $1/2$ 
but no isoscalar $NN$ component, 
whereas the $\kbar\kbar NN(1^+,1)$ state contains a pure isoscalar $NN$ component but
no isoscalar $\kbar N$ correlation with a fraction $1/4$ of the $T=0$ component. 
For these two states, ${\cal S}_\kbar$ is constant at
$9.1$~MeV for the set-I case and $19.5$~MeV for the set-II case.
The energy difference between  $\kbar\kbar NN(0^+,1)$ and $\kbar\kbar NN(1^+,1)$ is
the same value as the energy difference between 
$\kbar NN(0^-,1/2)$ and $\kbar NN(1^-,1/2)$, i.e., 
approximately $5$~MeV in the set-I result 
and $\approx 20$~MeV in the set-II result.

In future experimental searches for double-kaonic nuclei,  
the $\kbar\kbar NN$ states might be observed as the quasibound resonances 
in the invariant mass spectra of such modes as the $\Lambda\Lambda$, 
$\Lambda\Sigma^\pm\pi^\mp$, and $\Xi^- p$ decays.
The $\Lambda\Lambda$ mode for the $T=0$ spectrum shows the $\kbar\kbar NN(0^+,0)$ 
contribution, whereas the $\Lambda\Sigma^\pm\pi^\mp$ and $\Xi^- p$ modes probe 
both the $T=0$ and $T=1$ components and may contain the
$\kbar\kbar NN(0^+,1)$ and $\kbar\kbar NN(1+,1)$ contributions 
at higher energies than the $\kbar\kbar NN(0^+,0)$ contribution.

\subsection{Antiknock binding in single-kaonic nuclei}
\begin{figure}[!htp]
\includegraphics[width=7 cm]{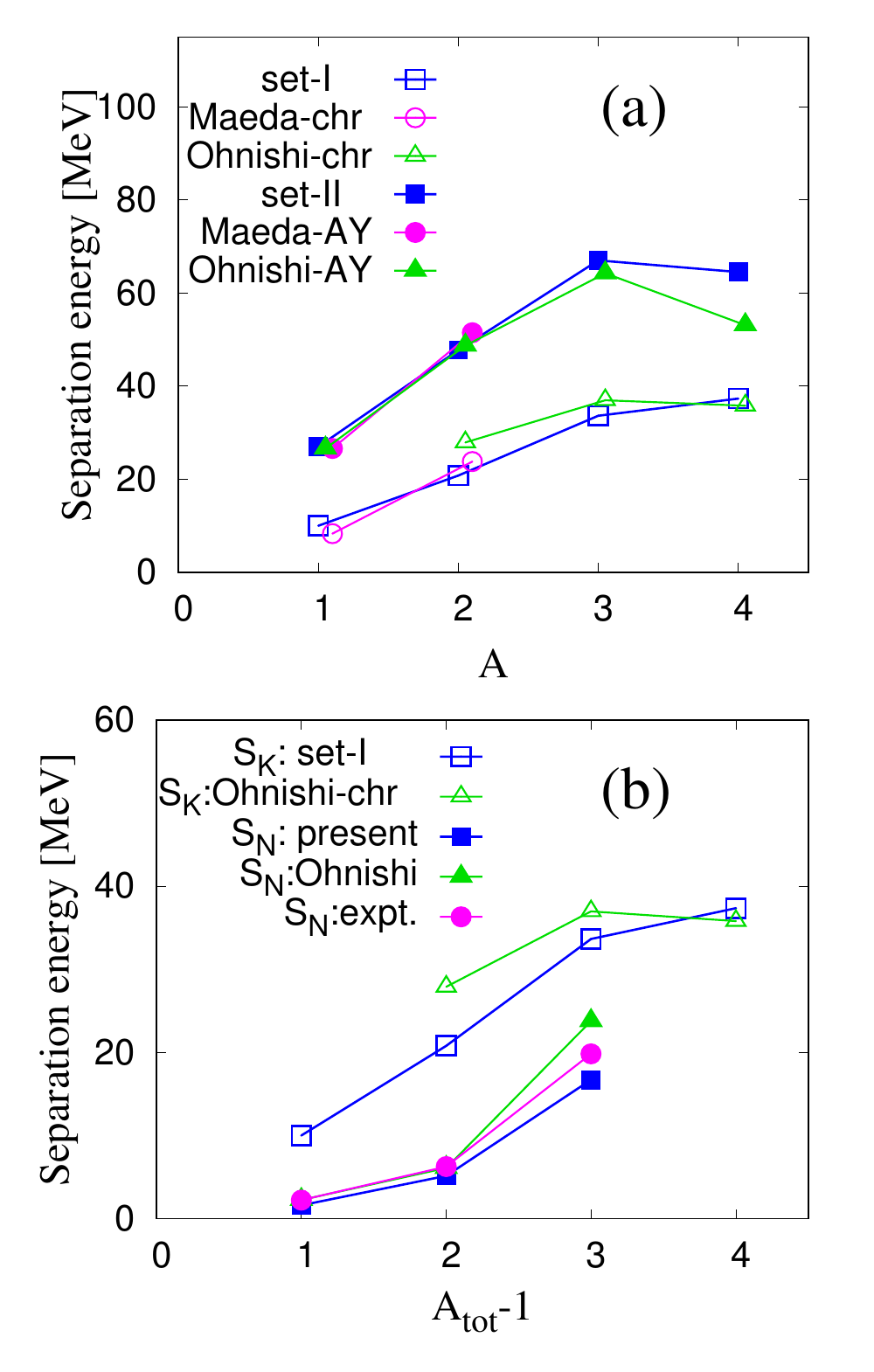}
\caption{$\kbar$-separation energies  (${\cal S}_\kbar$) of kaonic nuclei and nucleon-separation 
 energies  (${\cal S}_N$) of normal nuclei. 
(a) The set-I~(weak-binding) and set-II~(deep-binding) results of ${\cal S}_\kbar$
for the lowest states of single-kaonic nuclei.
(b) The set-I~(weak-binding) results of ${\cal S}_\kbar$ for single-kaonic nuclei and of ${\cal S}_N$ 
for normal nuclei.
For comparison, other theoretical results for ${\cal S}_\kbar$  that were 
calculated with weak-type chiral 
interactions by Maeda {\it et al.} 
(weak-chiral-regime case)~\cite{Maeda:2013zha} and 
Ohnishi {\it et al.} (Kyoto type-I case)~\cite{Ohnishi:2017uni} are also presented, 
together with those with deep-type AY interaction calculated 
by Maeda {\it et al.}~\cite{Maeda:2013zha} and  
Ohnishi {\it et al.}~\cite{Ohnishi:2017uni}.
For the ${\cal S}_N$ of normal nuclei,  the experimental values, and the AV4' result 
by Ohnishi {\it et al.}~\cite{Ohnishi:2017uni} are also shown.
	\label{fig:separation}}
\end{figure}

In Fig.~\ref{fig:separation}(a), I plot the $\kbar$-separation energies~(${\cal S}_\kbar$) 
for the lowest states of single-kaonic nuclei calculated with set-I and II.
For comparison, I also show theoretical results from Refs.~\cite{Maeda:2013zha,Ohnishi:2017uni}. 
In each group of weak- and deep-type calculations, the $A$ dependence of ${\cal S}_\kbar$ exhibits
a similar trend, in
that ${\cal S}_\kbar$ increases gradually up to $A=3$ and becomes saturated at $A=4$
because the isoscalar $\kbar N$ correlation vanishes in the nuclear-isospin-saturated system.
In Fig.~\ref{fig:separation}(b), 
I compare the separation energies, ${\cal S}_\kbar$ for single-kaonic nuclei and ${\cal S_N}$ for normal nuclei of set-I, which are plotted as functions of $A_\textrm{tot}-1$.
In the $A=2$ and 3 systems,  
a nucleon in a normal nucleus is rather weakly bound
because of the relatively weak $NN$ interaction compared with an antikaon that is deeply bound by the 
strong $\kbar N$ attraction in a kaonic nucleus.
However, at $A=4$ for $^4$He, ${\cal S_N}$ increases drastically. This is in contrast with 
the gradual change of ${\cal S_\kbar}$ with the increase of $A$ in kaonic nuclei. 

\section{Results of $\kNkN$-cluster model} \label{sec:results-2}

\subsection{Effective $\Lambda^*$-$\Lambda^*$ interaction}

To investigate the effective $\Lambda^*$-$\Lambda^*$ interaction,  
I apply the $\kNkN$-cluster model to the $\kbar\kbar NN$ system with total isospin $T=0$. 
As described in Sec.~\ref{sec:NK-NK-cluster}, I assume $0s$-orbital  configuration for each $\kbar N$ cluster
and consider the two-cluster wave function with a distance $R$.
In the cluster limit at a large distance $R$, each $\kbar N$ cluster forms an isoscalar $\kbar N$ bound state 
that corresponds to the $\Lambda^*$ state. 
In this asymptotic $\Lambda^*+\Lambda^*$ state, two $S^\pi=0^+$ and $1^-$ channels
are allowed because of Fermi statistics of $\Lambda^*$ particles, and their energies are 
degenerate at $R\to \infty$.

The mixing of $\tau=0$ and $\tau=1$ components 
with respect to the isospin $\tau_K=\tau_N=\tau$ of $NN$ and $\kbar\kbar$ pairs, which are 
coupled to total isospin $T=0$,   
is taken into account in each $S^\pi$ channel.

The $\kNkN$-cluster wave function with $\tau$ mixing 
can smoothly connect two limits; the shell model state at $R\to 0$ and the asymptotic $\Lambda^*+\Lambda^*$ state at $R\to \infty$. 
As the two $\Lambda^*$-clusters approach each other, 
the isospin rearrangement occurs through the isospin exchange between 
two clusters via the $\kbar N$ and $NN$ interactions. 

Because of the Bose and Fermi statistics,   
there are selection rules in the spatial symmetry of the $\kbar\kbar$ and $NN$ pairs
as follows. In the $(S^\pi T)=(0^+ 0)$ channel,
the $\kNkN$-cluster system 
is described by a linear combination of two isospin components:
the $\tau=1$ component with spatial-even $NN$ and $\kbar\kbar$ pairs, and 
the $\tau=0$ component with spatial-odd $NN$ and $\kbar\kbar$ pairs.
In the shell-model limit at $R\to 0$, 
the former component goes to the $0s$-orbital $\kbar\kbar NN(0^+,0)$ state, 
which is the lowest state in the $0s$-orbital model. 
The latter $\tau=0$ component is forbidden in the $0s$-orbital model space and instead  
goes to a $(0s)^2(0p)^2$  configuration with an antikaon and a nucleon excited into $0p$-orbits.
On the other hand, in the $(S^\pi T)=(1^-0)$ channel  with negative parity, 
 either one of $NN$ and $\kbar\kbar$ pairs is a spatial-odd state. 
The $\tau=0$ component contains  
a spatial-odd $\kbar\kbar$ pair, whereas the $\tau=1$ component has a 
spatial-odd $NN$ pair. In the shell-model limit at $R\to 0$, 
they become excited $(0s)^3(0p)$ states. In the  $\tau=0$ component, there's 
an antikaon excitation and in the $\tau=1$ component, 
there(s a nucleon excitation.

Such selection rules for $\kbar\kbar$ and $NN$ pairs
play important roles in the effective $\Lambda^*$-$\Lambda^*$ interaction, particularly
at short distances. 
In the asymptotic $\Lambda^*+\Lambda^*$ state at a large $R$, each channel of $S^\pi=0^+$ and $1^-$ 
contains $\tau=1$ and $\tau=0$ components with a ratio of 3:1.
On the other hand, in the shell-model limit at $R\to 0$, 
the $\tau$-mixing in the $S^\pi=0^+$ channel is equivalent to the mixing of 
the $(0s)^4$ and $(0s)^2(0p)^2$ configurations, and that in the $S^\pi=1^-$ channel
corresponds to the configuration mixing of the antikaon and nucleon excitations in the $(0s)^3(0p)$ configuration. 

\begin{figure}[!htp]
\includegraphics[width=7 cm]{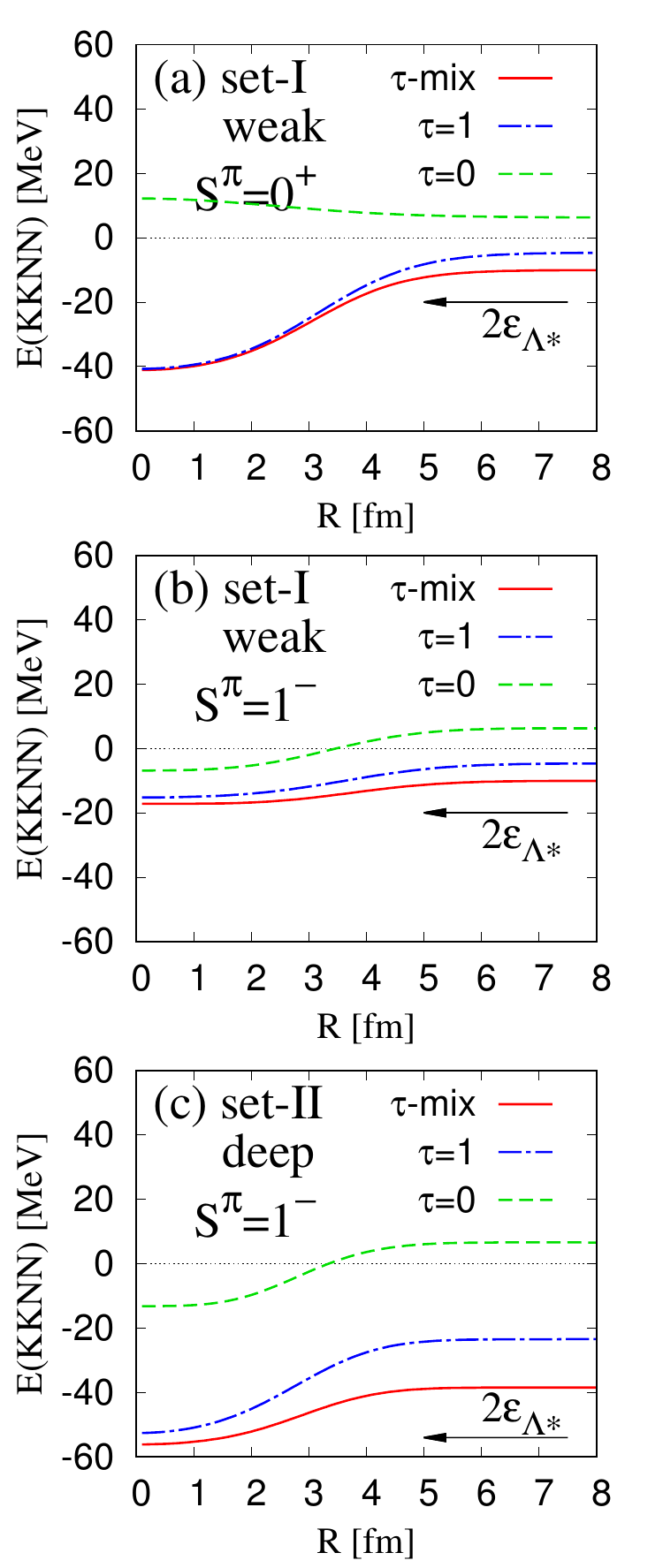}
\caption{
Energies of the $\kNkN$-cluster system with inter-cluster distances $R$
for (a) the $S^\pi=0^+$ and (b) $S^\pi=1^-$ states of the set-I result. 
(c) Those for the $S^\pi=1^-$ state of the set-II result. 
The energies calculated with and without $\tau$-mixing are shown. 
Arrows show the $\Lambda^*+\Lambda^*$ threshold energy, which is  
$\bar T_0$ below the asymptotic energy  at $R\to \infty$ obtained with $\tau$-mixing. 
	\label{fig:e-nknk}}
\end{figure}

I calculate the energy of the $\kNkN$-cluster state
with and without $\tau$-mixing at each distance $R$.
In Fig.~\ref{fig:e-nknk},
I show the $R$ dependence of the total energy of the $\kNkN$ state in the $S^\pi=0^+$ and $1^-$ channels. 
Note that 
the asymptotic $\Lambda^*+\Lambda^*$ state at a large distance $R$ 
contains an additional energy cost $\barT$ for localization of the relative motion, 
and the energy calculated with $\tau$-mixing equals $2\epsilon_{\Lambda^*}+\barT$ at sufficiently 
large $R$.
In both the $S^\pi=0^+$ and $1^-$ channels, 
I obtain the energy minimum at $R\to 0$ in the energy curve with $\tau$-mixing,
which indicates an attractive $\Lambda^*$-$\Lambda^*$ interaction. 
The attraction of the $\Lambda^*$-$\Lambda^*$ interaction in the $S^\pi=0^+$ channel 
is strong enough to form a deeply bound $\kbar\kbar NN(0^+,0)$ state in the shell-model limit, 
whereas that in the $S^\pi=1^-$ channel is weaker. 

In Table~\ref{tab:NK-NK}, I list values at $R\to 0$ 
for energy contributions and probability $P(\tau=1)$ for the $\tau=1$ component.
I also show the $\kbar$- and $\Lambda^*$-separation energies of the $\kbar\kbar NN$ states, which are
evaluated by the energy in the shell-model limit, as measured from 
the corresponding decay-threshold energies. 
For the $S^\pi=0^+$ state in the shell-model limit, the $\tau=1$ configuration is dominant and  
the $\tau=0$ mixing effect is negligibly small, meaning that the $0s$-orbital model used in the 
present work well approximates the deeply bound $\Lambda^*+\Lambda^*$ state 
in the $S^\pi=0^+$ channel. 

For the $S^\pi=1^-$ channel, I obtain a value of ${\cal S}_{\Lambda^*}=2.1$~MeV 
in the set-II result under $\tau$-mixing (see Table \ref{tab:NK-NK}), indicating that 
the $\Lambda^*$-$\Lambda^*$ attraction 
forms a (quasi) bound  $S^\pi=1^-$ state at a  slightly lower energy 
than the two-cluster-threshold energy, $2\epsilon_{\Lambda^*}$.
In the set-I result with $\tau$-mixing, a negative value ${\cal S}_{\Lambda^*}=-2.8$~MeV 
is obtained for the $\Lambda^*$-separation energy. 
This suggests that the $\Lambda^*$-$\Lambda^*$ attraction 
in the $S^\pi=1^-$ channel 
is insufficient to form a bound state, but may produce 
a $L^\pi=1^-$ resonance near the 2$\Lambda^*$-threshold energy.
The weaker $\Lambda^*$+$\Lambda^*$ attraction 
in the $S^\pi=1^-$ channel than in the $S^\pi=0^+$ channel 
is described by a much weaker $\kbar N$ attraction 
and a somewhat weaker $NN$ attraction as well as a larger kinetic-energy loss for the $0p$-orbit excitation.

Let me discuss the $S^\pi=1^-$ state in ore detail.
Comparing the energies of the $\tau=1$ and $\tau=0$ configurations without $\tau$-mixing, 
the $\tau=1$ component is favored  at all $R$ because of the stronger isoscalar $\kbar N$ correlation
than the $\tau=0$ component.
In the result with  $\tau$-mixing, the $\tau=1$ configuration dominates
the $S^\pi=1^-$ state in the shell-model limit with probability $P(\tau=1)=0.84$
for set-I and $P(\tau=1)=0.92$ for set-II.
In Table~\ref{tab:NK-NK-delta}, 
I show the energy contributions of the $\kNkN$ state at $R\to 0$, 
measured from twice of the internal-energy contributions of the $\Lambda^*$ cluster. From the table, 
one can see that the $\tau=1$ component with a single-nucleon excitation
gains energy in the $\kbar N$ attraction but loses energy due to $\kbar\kbar$ repulsion, whereas 
the $\tau=0$ component with a single-antikaon excitation gains energy through 
$NN$ attraction but somewhat 
loses energy through the $\kbar N$ attraction.

In principle, in the shell-model limit of $\kNkN$ with $S^\pi=1^-$, 
the $\tau=1$ and $\tau=0$ configurations 
can compete and the mixing ratio depends upon details of the interactions. 
In the present calculation, the nucleon excitation in the $\tau=1$ component is favored over
 the antikaon excitation in the $\tau=0$ component;
this can be explained by the 
nucleon feeling a broader $\kbar N$ mean-field, allowing it to more easily excite into the $0p$-orbit than 
an antikaon, because light-mass antikaons have
broader density distributions than nucleons in the present model.
The $\tau$-mixing ratio may change if the antikaon mass is heavier than the physical antikaon mass.
For example, if I assume equal kaon and nucleon masses $m_\kbar=m_N$ and keep the other parameters 
unchanged, 
I obtain a lower energy for the $\tau=0$ component with an antikaon excitation 
than for the $\tau=1$ component with a nucleon excitation in the small-$R$ region, 
resulting in significant $\tau$-mixing in the $\kNkN$ state with $S^\pi=1^-$. 

\subsection{Comparison of $\Lambda^*$-$\Lambda^*$ and $d$-$d$ interactions}
The $\Lambda^*$-$\Lambda^*$ interaction discussed previously is regarded as an effective 
dimer-dimer interaction in the kaonic nuclei. 
I here compare its properties with those of the $d$-$d$ interaction in nuclear systems.
A deuteron is a weakly bound $(ST)=(10)$ $NN$ state. I describe the $NN+NN$ system in the $S^\pi=0^+$, $1^-$, and $2^+$ channels with a $d+d$ cluster model called the Brink-Bloch model~\cite{brink66} as done in Ref.~\cite{Kanada-Enyo:2020zzf}. 
The detailed properties of the effective $d$-$d$ interaction
have been investigated in the previous paper~\cite{Kanada-Enyo:2020zzf}. In the present paper, I show the energy of the $NN+NN$ system at $R\to 0$ for the set-I parametrization,
and discuss the roles of kinetic- and potential-energy contributions in the effective dimer-dimer interactions
of the two systems. 

I can classify the $S^\pi$ states of two systems
based on the number of spatial-odd $\kbar\kbar$ and $NN$ pairs. 
Because of the nucleon Fermi statistics, 
the $S^\pi=0^+$, $S^\pi=1^-$, and $S^\pi=2^+$ states of the  
$NN+NN$ system contain zero, one, and two spatial-odd $NN$ pairs, respectively.
Similarly, in the $\kbar N+\kbar N$ system, the $\tau=1$ and  $\tau=0$ components of the 
$S^\pi=0^+$ state contain zero and two spatial-odd pairs, respectively, while the $S^\pi=1^-$ state
has one spatial-odd pair.
In the shell-model limit, these states having no, one, and two spatial-odd pairs correspond to the 
$(0s)^4$, $(0s)^3(0p)$,  $(0s)^2(0p)^2$ configurations, which have 
kinetic energy of 
$3\barT$, $(3+\frac{2}{3})\barT$, and $(3+\frac{4}{3})\barT$, respectively.

\begin{figure}[!htp]
\includegraphics[width=8 cm]{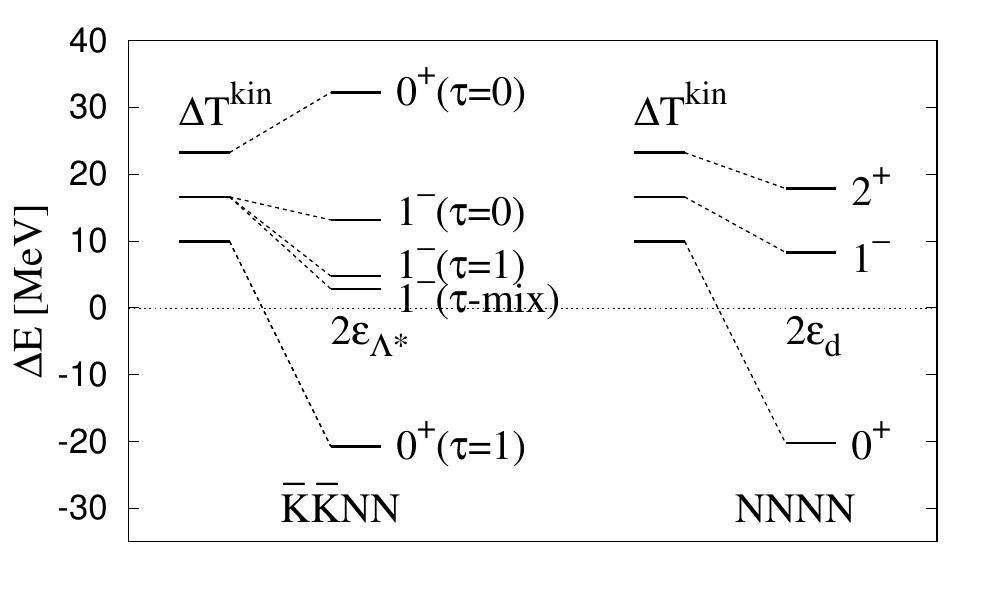}
\caption{Energy spectra of the $\kbar\kbar NN$ and $NNNN$ systems calculated with the 
$\kNkN$ and $NN+NN$-cluster models in the shell-model limit. 
The energies are measured from the 
two-cluster threshold energies, $2\epsilon_{\Lambda^*}$ for the $\kbar\kbar NN$ system 
and $2\epsilon_{d}$ for the $NNNN$ system. 
For the $\kbar\kbar NN$ system, the energies without  $\tau$-mixing 
and the $S^\pi=1^-$ energy with $\tau$-mixing
are shown.
Energy levels above the threshold are not bound states, but energies obtained 
for the lowest shell-model configurations are plotted. 
	\label{fig:d-d-spe}}
\end{figure}

In Fig.~\ref{fig:d-d-spe} and 
Table \ref{tab:NK-NK-delta},  
I show the results of the $\kNkN$ and $NN+NN$ systems in the shell-model limit.
The energy contributions measured from 
twice of the internal energies of a single cluster are shown.
From the energy spectra of Fig.~\ref{fig:d-d-spe},
the two clusters in the lowest $S^\pi=0^+$ channel for the $(0s)^4$ configuration 
are deeply bound in both systems.
The binding energy for the two $\Lambda^*$ clusters from the threshold is approximately 20 MeV, 
coinciding with that for two deuteron clusters. However, 
the detailed contributions of the four pairs between the two clusters differ.
According to the energy counting in the present model, 
the $\Lambda^*+\Lambda^*$ and $d+d$ binding energies are given as 
\begin{align}
\Delta E^{(0^+,0)}_{\kbar \kbar NN}&\equiv E^{(0^+,0)}_{\kbar\kbar NN}-2\epsilon_{\Lambda^*}
\nonumber\\
&=\barT+\barV^{(01)}_{NN}
+\barV^{T=0}_{\kbar N}+\barV^{T=1}_{\kbar N}
+\barV^{T=1}_{\kbar\kbar},\\
\Delta E^{(0^+,0)}_{NNNN}&\equiv E^{(0^+,0)}_{NNNN}-2\epsilon_{d}
\nonumber\\
&=\barT+3\barV^{(01)}_{NN}+\barV^{(10)}_{NN}.
\end{align}
In the $NNNN$ system,
the potential energy contribution is always attractive for all four $NN$ pairs between the two clusters, 
while in the $\kbar\kbar NN$ system, the strong attraction in the isocalar $\kbar N$ pair compensates for
the repulsion in the isovector $\kbar\kbar$ pair.

In the $S^\pi=1^-$ channel of the $NNNN$ and $\kbar\kbar NN$ systems at $R\to 0$, 
two clusters gain some amount of potential energy but 
lose kinetic energy for one $0p$-orbit excitation in the $(0s)^3(0p)$ configuration.
In the $\kbar\kbar NN$ system, 
the $\Lambda^*$-$\Lambda^*$ interaction in the $S^\pi=1^-$ channel
is the weak attraction and almost forms a bound state at an energy close to 
the $2\Lambda^*$ threshold.
To gain the $\kbar N$ attraction efficiently from the 
$\Lambda^*+\Lambda^*$ state to the shell-model-limit state, 
isospin rearrangement plays an essential role. This is 
a unique characteristic of kaonic nuclei, but cannot be seen in the $NN+NN$ system
because such isospin rearrangement is not allowed in the $(S^\pi T)=(1^- 0)$ state. Hence,
there is no attraction of the $d$-$d$ interaction in the $S^\pi=1^-$ channel.

The $NNNN(2^+0)$ state and the $\tau=0$ component of the $\kbar\kbar NN(0^+0)$ state 
correspond to the $(0s)^2(0p)^2$ configuration and have 
much higher energy than the two-cluster threshold. 
A comparison of the two systems shows that
the kinetic-energy loss is the same, but 
the total energy differs significantly  
because of the difference in the potential-energy contributions 
(see Fig.~\ref{fig:d-d-spe}).
As shown in Table~\ref{tab:NK-NK-delta}, 
the $NNNN(2^+0)$ state gains potential energy because of the attractive $NN$ interaction, 
whereas  the $\tau=0$ component of the $\kbar\kbar NN(0^+ 0)$ state containing 
isoscalar $\kbar\kbar$ and $NN$ pairs  
looses the potential energy of the $\kbar N$ and $\kbar\kbar$ interactions. 

\begin{table*}[!ht]
\caption{Energies of the $\kNkN(S^\pi T)$ states in the shell-model $(R\to 0)$ limit, as calculated 
with and without $\tau$-mixing. The set-I and II results are shown in the 
upper and lower parts, respectively. 
\label{tab:NK-NK}}
\begin{center}
\begin{tabular}{cccccccccc}
\hline
\multicolumn{8}{l}{set-I ($\epsilon_{\Lambda^*}=-10$~MeV, $\nu_N=0.16$~fm$^{-2}$)}\\																	
	&	$\langle  T^\textrm{kin} \rangle$	&	$\langle  v_{NN} \rangle$	&	$\langle  v_{\kbar N} \rangle$	&	$\langle  v_{\kbar\kbar} \rangle$	&	$E$	&	${\cal S}_{\kbar}$	&	${\cal S}_{\Lambda^*}$	&	$P(\tau=1)$	\\
\multicolumn{4}{l}{$\kbar N+\kbar N\ \ (S^\pi T)=(0^+ 0)$}\\																	
${\tau}$-mixing	&	30.0 	&$	-6.2 	$&$	-69.5 	$&$	4.6 	$&$	-41.1 	$&$	20.3 	$&$	21.1 	$&	0.99 	\\
$\tau=1$	&	29.9 	&$	-6.2 	$&$	-69.0 	$&$	4.6 	$&$	-40.7 	$&$	19.9 	$&$	20.7 	$&	1	\\
$\tau=0$	&	43.1 	&$	0.7 	$&$	-31.7 	$&$	0.0 	$&$	12.2 	$&$	-33.0 	$&$	-32.2 	$&	0	\\
\multicolumn{4}{l}{$\kbar N+\kbar N\ \ (S^\pi T)=(1^- 0)$}\\																	
${\tau}$-mixing	&	36.5 	&$	-1.2 	$&$	-56.3 	$&$	3.9 	$&$	-17.2 	$&$	1.9 	$&$	-2.8 	$&	0.84 	\\
$\tau=1$	&	36.5 	&$	0.7 	$&$	-57.1 	$&$	4.6 	$&$	-15.2 	$&$	-0.1 	$&$	-4.8 	$&	1	\\
$\tau=0$	&	36.5 	&$	-11.6 	$&$	-31.8 	$&$	0.0 	$&$	-6.9 	$&$	-8.4 	$&$	-13.1 	$&	0	\\
&\\																	
\multicolumn{8}{l}{set-II ($\epsilon_{\Lambda^*}=-27$~MeV, $\nu_N=0.25$~fm$^{-2}$)}\\																	
	&	$\langle  T^\textrm{kin} \rangle$	&	$\langle  v_{NN} \rangle$	&	$\langle  v_{\kbar N} \rangle$	&	$\langle  v_{\kbar\kbar} \rangle$	&	$E$	&	${\cal S}_{\kbar}$	&	${\cal S}_{\Lambda^*}$	&	$P(\tau=1)$	\\
\multicolumn{4}{l}{$\kbar N+\kbar N\ \ (S^\pi T)=(0^+ 0)$}\\																	
${\tau}$-mixing	&	46.9 	&$	-8.7 	$&$	-141.8 	$&$	4.9 	$&$	-98.7 	$&$		$&$		$&	0.99 	\\
$\tau=1$	&	46.7 	&$	-8.8 	$&$	-140.2 	$&$	5.0 	$&$	-97.3 	$&$	49.6 	$&$	43.3 	$&	1	\\
$\tau=0$	&	67.4 	&$	1.4 	$&$	-53.6 	$&$	0.0 	$&$	15.2 	$&$		$&$		$&	0	\\
\multicolumn{4}{l}{$\kbar N+\kbar N\ \ (S^\pi T)=(1^- 0)$}\\																	
${\tau}$-mixing	&	57.0 	&$	0.0 	$&$	-117.7 	$&$	4.6 	$&$	-56.1 	$&$	30.9 	$&$	2.1 	$&	0.92 	\\
$\tau=1$	&	57.0 	&$	1.4 	$&$	-116.0 	$&$	5.0 	$&$	-52.6 	$&$	27.3 	$&$	-1.4 	$&	1	\\
$\tau=0$	&	57.0 	&$	-16.3 	$&$	-53.9 	$&$	0.0 	$&$	-13.2 	$&$	-12.1 	$&$	-40.8 	$&	0	\\
\hline
\end{tabular}
\end{center}
\end{table*}

\begin{table*}[!ht]
\caption{Energies of the $\kNkN(S^\pi T)$ and $NN+NN(S^\pi T)$ states 
in the shell-model limit, as 
measured from the two-cluster threshold energies. 
The set-I result of the 
total energy ($\Delta E$), kinetic~($\Delta T^\textrm{kin}$), $NN$~($\Delta v_{NN}$), $\kbar N$~($\Delta v_{\kbar N}$), and $\kbar\kbar$~($\Delta v_{\kbar\kbar}$) interaction-energy contributions 
measured from twice of the internal energies of a single cluster are listed. 
For the $\kNkN$ states, the results obtained with and without $\tau$-mixing are shown. 
All energies are in units of MeV.
 \label{tab:NK-NK-delta}}
\begin{center}
\begin{tabular}{cccccccccc}
\hline
\multicolumn{5}{l}{set-I ($\epsilon_{\Lambda^*}=-10$~MeV, $\nu_N=0.16$~fm$^{-2}$)}\\
	&	configuration	&	$  \Delta T^\textrm{kin}$	&	$ \Delta v_{NN} $	&	$ \Delta v_{\kbar N} $	&	$ \Delta v_{\kbar\kbar} $	&	$ \Delta E $	\\	
\multicolumn{4}{l}{$\kbar N+\kbar N~(0^+ 0)$}\\													
${\tau}$-mixing	&	$(0s)^4+(0s)^2(0p)^2$	&	10.0 	&$	-6.2 	$&$	-29.6 	$&$	4.6 	$&$	-21.1 	$\\
$\tau=1$	&	$(0s)^4$	&	10.0 	&$	-6.2 	$&$	-29.1 	$&$	4.6 	$&$	-20.7 	$\\
$\tau=0$	&	$(0s)^2(0p)^2$	&	23.2 	&$	0.7 	$&$	8.2 	$&$	0.0 	$&$	32.2 	$\\
\multicolumn{4}{l}{$\kbar N+\kbar N~(1^- 0)$}\\													
${\tau}$-mixing	&	$(0s)^3(0p)^1$	&	16.6 	&$	-1.2 	$&$	-16.4 	$&$	3.9 	$&$	2.8 	$\\
$\tau=1$	&	$(0s)^3(0p)^1$	&	16.6 	&$	0.7 	$&$	-17.2 	$&$	4.6 	$&$	4.8 	$\\
$\tau=0$	&	$(0s)^3(0p)^1$	&	16.6 	&$	-11.6 	$&$	8.1 	$&$	0.0 	$&$	13.1 	$\\
&\\													
	&		&	$ \Delta T^\textrm{kin}$	&	$ \Delta v_{NN} $	&		&		&	$ \Delta E$	\\
\multicolumn{4}{l}{$NN+NN~(0^+ 0)$}\\													
	&	$(0s)^4$	&	10.0 	&$	-30.2 	$&$		$&$		$&$	-20.3 	$\\
\multicolumn{4}{l}{$NN+NN~(1^- 0)$}\\													
	&	$(0s)^3(0p)^1$	&	16.6 	&$	-8.3 	$&$		$&$		$&$	8.3 	$\\
\multicolumn{4}{l}{$NN+NN~(2^+ 0)$}\\													
	&	$(0s)^2(0p)^2$	&	23.2 	&$	-5.4 	$&$		$&$		$&$	17.8 	$\\
\hline
\end{tabular}
\end{center}
\end{table*}

\section{Summary} \label{sec:summary}

I investigated the energy systematics of single- and double-kaonic nuclei in the mass number $A\le 4$ region
with the $0s$-orbital model using 
zero-range $\kbar N$ and $\kbar\kbar$ interactions. 
The $\kbar N$ interaction was tuned to fit the $\Lambda(1405)$ mass with the energy of the $\kbar N$ bound state. 
For the $NN$ interaction, I adopted the Volkov finite-range central interaction with a tuned 
parametrization adjusted to reproduce the $S$-wave $NN$-scattering lengths.  
I calculated the energy spectra of the $\kbar NN$, $\kbar NNNN$, $\kbar NNN$, $\kbar\kbar N$,
and $\kbar\kbar NN$ systems in the cases of weak- and deep-binding and compared the results with 
other theoretical calculations with weak-type chiral and deep-type AY interactions.
The present results qualitatively reproduce the energy systematics of kaonic nuclei calculated via
other theoretical approaches. In the present $0s$-orbital model, 
the energy spectra of kaonic nuclei were given by simple energy counting 
of isospin components of $NN$, $\kbar N$, $\kbar\kbar$ pairs. 
The approximate energy relations for the lowest states of the $\kbar N$,  $\kbar NN$, and  $\kbar\kbar NN$ systems were obtained
as $E_{\kbar NN}^{(0^-,1/2)}\approx 2\epsilon_{\Lambda^*}$ and 
$E_{\kbar\kbar NN}^{(0^+,0)}\approx 2E_{\kbar NN}^{(0^-,1/2)}$, which are universal features that are 
independent of the $\Lambda(1405)$ mass. 

For the $\kbar NN$ and $\kbar\kbar NN$ systems, 
I discussed the important roles of the isospin symmetry
in the energy spectra of the $(J^\pi,T)$ states.
In addition to the 
lowest $\kbar NN(0^-,1/2)$ and $\kbar\kbar NN(0^+,0)$ states containing the isovector $(ST)=(01)$ $NN$ pair, 
I also 
obtain the $\kbar\kbar NN(0^+,1)$,
$\kbar NN(1^-,1/2)$, and $\kbar\kbar NN(1^+,1)$ states. 
The latter two have isoscalar $(ST)=(10)$ $NN$ pairs like deuterons.
The predicted $\kbar$-separation energies for these states 
are ${\cal S}_\kbar$=9--25 MeV.
In future experimental searches for $\kbar\kbar NN$ states, 
the $\kbar\kbar NN(0^+,1)$ and  $\kbar\kbar NN(1^+,1)$ states may contribute to the $T=1$ components of 
invariant mass spectra.

I also investigated the effective $\Lambda^*$-$\Lambda^*$ interaction with the 
$\kNkN$-cluster model and obtained a strong attraction in the $S^\pi=0^+$ channel and 
a weak attraction in the $S^\pi=1^-$ channel. 
In comparing the $\Lambda^*$-$\Lambda^*$ interaction in the $\kbar\kbar NN$ system 
with the $d$-$d$ interaction in the $NNNN$ system, 
I discussed  the properties of dimer-dimer interactions in hadron and nuclear systems. 

In the present calculation, zero-range 
real potentials were used for the $\kbar N$ and $\kbar\kbar$ interactions. Moreover, kaonic nuclei were simply 
described using the $0s$-orbital and cluster models.
Despite such simple theoretical treatments of interactions and wave functions, 
the present results succeeded in globally describing the energy systematics of kaonic nuclei 
obtained with precise few-body calculations.
The energy-counting rule in the
present model is useful for  understanding the leading properties of energy spectra in kaonic nuclei; 
it also 
enables one to extract universal features independently from the details of the hadron-hadron interactions. 
For precise predictions of the energies and widths of the quasibound states of kaonic nuclei, 
it is necessary to perform further investigations with sophisticated 
calculations beyond the present framework.
Higher-order effects in hadron interactions such as the 
 $\kbar N$-$\pi\Sigma$ coupling or imaginary part, the 
energy dependences of the $\kbar N$ interaction, and  
non-central $NN$ forces should be also taken into account.

\begin{acknowledgments}  
This work was inspired by discussions with Prof.~Lee for dimer-dimer interactions.
The numerical calculations of this work were performed using the
computer systems at the Yukawa Institute for Theoretical Physics at Kyoto University. The work was supported
by Grants-in-Aid of the Japan Society for the Promotion of Science (Grant Nos. JP18K03617 and 18H05407).
\end{acknowledgments}



\end{document}